\documentclass[]{aa}  

\usepackage{hyperref}
\hypersetup{colorlinks, allcolors=blue}
\usepackage{graphicx}    
\usepackage{color}  
\usepackage{txfonts}
\usepackage{inputenc}
\usepackage{multicol}
\usepackage{multirow}
\usepackage{siunitx}
\DeclareSIUnit{\solarmass}{\text{M}_{\odot}}
\DeclareSIUnit{\parsec}{pc}
\DeclareSIUnit{\year}{yr}
\DeclareSIUnit{\mag}{mag}
\DeclareSIUnit{\arcsec}{arcsec}
\usepackage{amsmath}
\usepackage{xcolor}
\usepackage{booktabs}
\usepackage{color} 

\usepackage{algorithm} 
\usepackage{algpseudocode} 

\makeatletter
\renewcommand*\aa@pageof{, page \thepage{} of \pageref*{LastPage}}
\makeatother

\begin{document}

\title{The star formation history of the Sco-Cen association}
    \subtitle{Coherent star formation patterns in space and time}
   
   \author{Sebastian Ratzenböck\inst{1,2,3} 
          \and
          Josefa~E.~Gro\ss schedl\inst{1}
          \and
          Jo\~ao Alves\inst{1,2} 
          \and
          N\'uria Miret-Roig\inst{1}
          \and
          Immanuel Bomze\inst{2,4}
          \and
          John Forbes\inst{5,6}
          \and
          Alyssa Goodman\inst{7}
          \and
          \'Alvaro Hacar\inst{1}
          \and
          Doug Lin\inst{8,9}
          \and
          Stefan Meingast\inst{1}
          \and
          Torsten M\"oller\inst{2,3}
          \and
          Martin Piecka\inst{1} 
          \and
          Laura Posch\inst{1} 
          \and
          Alena Rottensteiner\inst{1}
          \and
          Cameren Swiggum\inst{1}
          \and
          Catherine Zucker\inst{10}
          }

   \institute{University of Vienna, Department of Astrophysics,
              T\"urkenschanzstra{\ss}e 17, 1180 Vienna, Austria\\
              \email{sebastian.ratzenboeck@univie.ac.at}
         \and
             University of Vienna, Data Science at Uni Vienna Research Platform, Austria
        \and 
            University of Vienna, Faculty of Computer Science, W\"ahringer Straße 29/S6, A-1090 Vienna
        \and
           University of Vienna, ISOR/VCOR, Oskar-Morgenstern-Platz 1, A-1090 Vienna    
        \and
           School of Physical and Chemical Sciences - Te Kura Matū, University of Canterbury, Christchurch 9150, New Zealand
         \and  
           Center for Computational Astrophysics, Flatiron Institute, 162 5th Avenue New York, NY, 10010, USA
        \and
           Harvard-Smithsonian Center for Astrophysics
            60 Garden Street, MS- 42
            Cambridge, MA 02138, USA  
        \and
        Department of Astronomy and Astrophysics, University of California, Santa Cruz, CA, USA
        \and
        Institute for Advanced Studies, Tsinghua University, Beijing, People’s Republic of China
        \and
            Space Telescope Science Institute, 3700 San Martin Drive, Baltimore, MD 21218, USA
             }
             
   \date{Received ... ; accepted ... }
   
  \abstract
    {
    We reconstruct the star formation history of the Sco-Cen OB association using a novel high-resolution age map of the region. We develop an approach to produce robust ages for Sco-Cen's recently identified 37 stellar clusters using the \texttt{SigMA} algorithm. The Sco-Cen star formation timeline reveals four periods of enhanced star formation activity, or bursts, remarkably separated by about 5\,Myr. Of these, the second burst, which occurred about 15 million years ago, is by far the dominant, and most of Sco-Cen's stars and clusters were in place by the end of this burst. The formation of stars and clusters in Sco-Cen is correlated but not linearly, implying that more stars were formed per cluster during the peak of the star formation rate. Most of the clusters that are large enough to have supernova precursors were formed during the 15\,Myr period. Star and cluster formation activity has been continuously declining since then. We have clear evidence that Sco-Cen formed from the inside out and contains 100-pc long chains of contiguous clusters exhibiting well-defined age gradients, from massive older clusters to smaller young clusters. These observables suggest an important role for feedback in forming about half of Sco-Cen stars, although follow-up work is needed to quantify this statement. Finally, we confirm that the Upper-Sco age controversy discussed in the literature during the last decades is solved: the nine clusters previously lumped together as Upper-Sco, a benchmark region for planet formation studies, exhibit a wide range of ages from 3 to 19\,Myr.   
    }


   \maketitle
\defcitealias{Kerr2021}{KRK21}
\defcitealias{Schmitt2022}{SCF22}
\defcitealias{Squicciarini2021}{SGB21}
\defcitealias{Fang2017}{F17}
\defcitealias{Ratzenboeck2022}{Paper\,I}
\defcitealias{Baraffe2015}{BHAC15}

\section{Introduction}\label{sec:intro}

Reconstructing the star formation history of a star-forming region is essential for gaining insight into the complex and out-of-equilibrium process of star formation. A timeline of when and where different sub-populations form offers insights into the underlying physical mechanisms driving star formation. For example, is the star formation rate in a collapsing cloud accelerating until gas consumption? Does star formation leave behind discernible spatio-temporal patterns? Or is it a chaotic process? And if it leaves patterns of progression, are they intrinsic to the process or driven by an external agent? Can variations in the star formation rate directly relate to fundamental properties of the resulting stellar population, for instance, the formation of gravitationally bound clusters as opposed to loose stellar associations? Star formation history encodes much of the information needed to address these questions. It is crucial for developing accurate star formation models and interpreting observations of star-forming regions across the universe.

Unfortunately, recognizing distinct populations in a star-forming region is challenging, particularly for loose stellar associations that quickly disperse into the surrounding Galactic field, making them difficult to identify. Moreover, as they are formed from the same molecular cloud complex, the velocities and age differences between different sub-populations are small and, therefore, hard to measure. Despite the growing evidence that sub-populations exist inside the same star formation region \citep[e.g.,][]{Alves2012, Jerabkova2019, Chen2020, Grossschedl2021, Kerr2021, Luhman2022a, MiretRoig2022b}, there is little evidence so far for large-scale star formation patterns \citep{Wright2022arXiv}.

The Scorpius-Centaurus OB association (Sco-Cen) \citep{Blaauw1946, Blaauw1964a} is the closest OB association to Earth. The full association is a large, roughly 200-pc-wide complex that still includes molecular clouds with ongoing star formation. It is an ideal laboratory for studying various aspects of star, planet, and stellar cluster formation and evolution. It was well established early on that Sco-Cen includes sub-populations with ages from about 1\,Myr to about 20\,Myr. 
\citep[e.g.,][]{DeGeus1989, DeGeus1992, DeBruijne1999, Preibisch1999, deZeeuw1999, DeZeeuw2001, Lepine2003, Preibisch2008, Makarov2007a, Makarov2007b, Diehl2010, Poeppel2010, Rizzuto2011, Pecaut2012, Pecaut2016, Krause2018, Forbes2021}.  The availability of \textit{Gaia} data \citep{Brown2016} has generated a series of Sco-Cen studies that aim to determine the structure of the association with the superb astrometric and photometric data \citep[e.g.,][]{Wright2018, VillaVelez2018, Goldman2018, Damiani2019, LuhmanEsplin2020, Grasser2021, Kerr2021, Squicciarini2021, Schmitt2022, Luhman2022a, MiretRoig2022b, Ratzenboeck2022, BricenoMorales2023}.

Recently, \cite{Ratzenboeck2022} presented results from a novel clustering algorithm called Significance Mode Analysis (\texttt{SigMA}), which interprets density peaks separated by dips as significant clusters. Using a graph-based approach, the technique detects peaks and dips directly in the 5D multidimensional phase space. The method can identify co-spatial and co-moving clusters with non-convex shapes and variable densities with a measure of significance. The application of \texttt{SigMA} to \textit{Gaia} DR3 data \citep{Vallenari2022arXiv} of stars in and around the Sco-Cen association led to the discovery of multiple clusters\footnote{Henceforth, we use the word ``cluster'' in the statistical sense, namely, an enhancement over a background, as extracted with \texttt{SigMA}. This avoids creating a new word for the spatial/kinematical coherent structures in Sco-Cen. None of the Sco-Cen clusters is expected to be gravitationally bound.}, reaching stellar volume densities as low as 0.01\,sources/pc$^3$ and tangential velocity differences of about 0.5\,km\,s$^{-1}$ between different clusters. This level of accuracy is unprecedented and has unveiled 37 stellar clusters inside Sco-Cen. The \texttt{SigMA} algorithm opens new possibilities for a detailed look at the star formation history of Sco-Cen and other nearby star formation complexes.

The goal of this paper is to derive robust ages for the 37 stellar Sco-Cen clusters identified by \citet{Ratzenboeck2022} with the \texttt{SigMA} algorithm (hereafter, \citetalias{Ratzenboeck2022}) and reconstruct the Star Formation History of this important region. Given \texttt{SigMA}'s ability to disentangle young populations in Sco-Cen, adding robust ages to these clusters should enable the construction of a high-resolution age map of the region. The paper is structured as follows. In Sect.~\ref{sec:data}, we present the data and the quality criteria applied. In Sect.~\ref{sec:method} we roughly summarize our methods to determine robust isochronal ages, while we outline the methods in detail in Appendix~\ref{apx:method}. In Sect.~\ref{sec:resutls}, we present  our results, which we then discuss in Sect.~\ref{sec:discussion}. In Sect.~\ref{sec:conclusion}, we summarize our conclusions.

\section{Data} \label{sec:data}

\begin{figure*}[!t]
    \centering
    \includegraphics[width=0.9\textwidth]{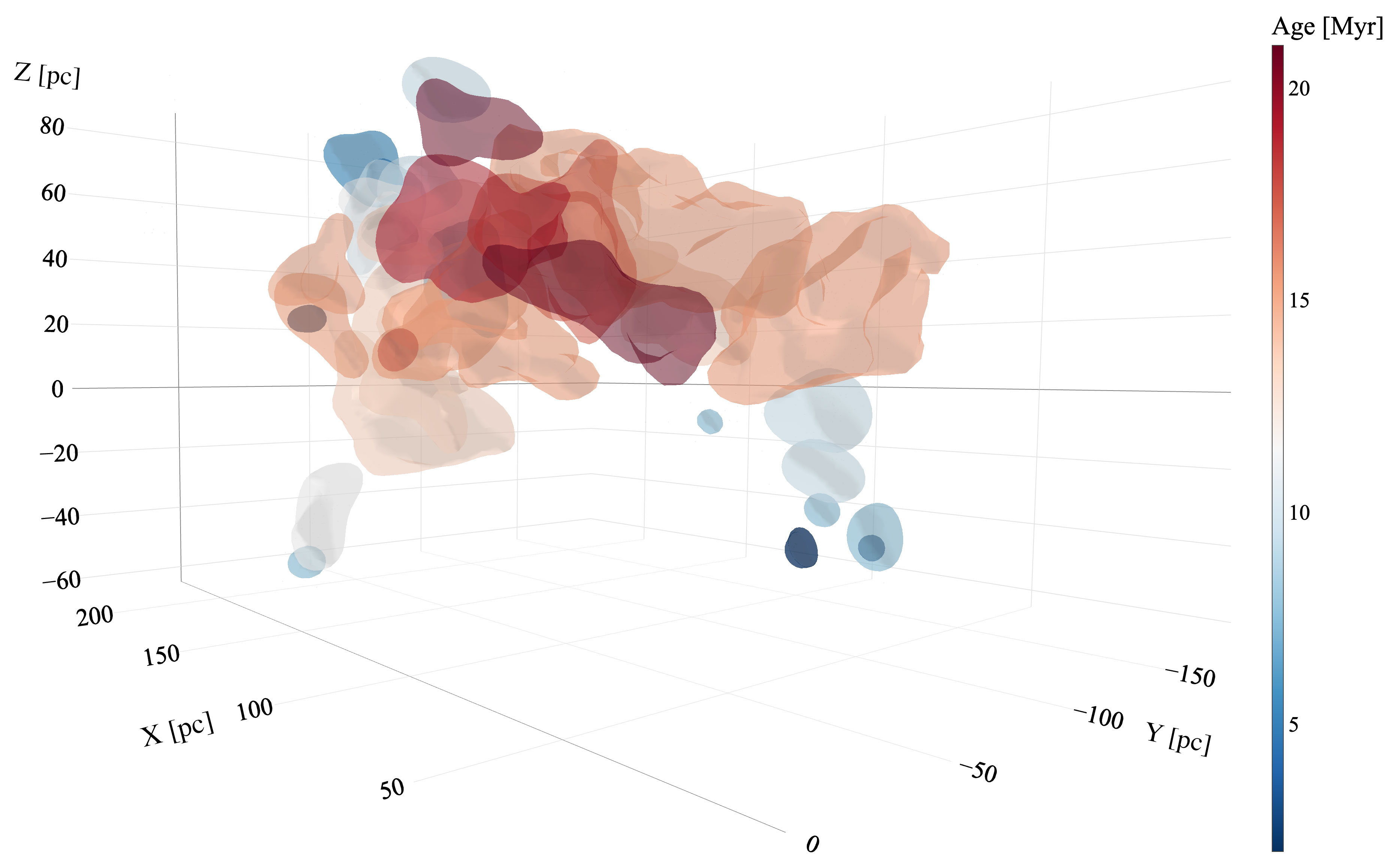}
    \caption{3D distribution of 34 clusters in the Sco-Cen association found by \texttt{SigMA}. The Sun is at (0,0,0) and the Z=0 plane is parallel to the Galactic plane. The surfaces of the cluster volumes are shown, color-coded by age, from dark blue (2 Myr) to dark red (21 Myr). For more details, see the link to the \href{https://homepage.univie.ac.at/sebastian.ratzenboeck/wp-content/uploads/2023/05/scocen_age.html}{interactive 3D version}.
    } 
    \label{fig:ScoCen3D-ages}
\end{figure*}

\begin{table*}
\begin{center}
\begin{small}
\caption{Isochronal ages for the 37 stellar groups in Sco-Cen as selected with \texttt{SigMA} in \citet{Ratzenboeck2022}.\vspace{-1,5mm}}
\renewcommand{\arraystretch}{1.4}
\resizebox{0.9\textwidth}{!}{
\begin{tabular}{lllrrrrrrrr}

\hline \hline
\multicolumn{1}{l}{\texttt{SigMA}} &
\multicolumn{1}{l}{Traditional} &
\multicolumn{1}{l}{Group Name} &
\multicolumn{3}{c}{Number of sources\tablefootmark{a}} &
\multicolumn{2}{c}{Age PARSEC} &
\multicolumn{2}{c}{Age BHAC15} \\
[-0.9mm] 

\cmidrule(lr){4-6}
\cmidrule(lr){7-8}
\cmidrule(lr){9-10}

\multicolumn{1}{l}{} &
\multicolumn{1}{l}{Regions} &
\multicolumn{1}{l}{} &
\multicolumn{1}{c}{All} &
\multicolumn{1}{c}{BPRP} &
\multicolumn{1}{c}{GRP} &
\multicolumn{1}{c}{BPRP} &
\multicolumn{1}{c}{GRP} &
\multicolumn{1}{c}{BPRP} &
\multicolumn{1}{c}{GRP} \\
[-0.9mm] 
\multicolumn{1}{l}{} &
\multicolumn{1}{l}{} &
\multicolumn{1}{l}{} &
\multicolumn{1}{l}{} &
\multicolumn{1}{l}{} &
\multicolumn{1}{l}{} &
\multicolumn{1}{c}{(Myr)} &
\multicolumn{1}{c}{(Myr)} &
\multicolumn{1}{c}{(Myr)} &
\multicolumn{1}{c}{(Myr)} \\

\hline

1 & US & $\rho$ Oph/L1688 & 535 & 338 & 350 & \phantom{0}3.8 (\phantom{0}3.4, \phantom{0}4.2) & \phantom{0}3.6 (\phantom{0}3.3, \phantom{0}4.1) & \phantom{0}3.2 (\phantom{0}3.1, \phantom{0}3.4) & \phantom{0}4.6 (\phantom{0}4.4, \phantom{0}4.8) \\
2 & US & $\nu$ Sco & 150 & 98 & 101 & \phantom{0}5.8 (\phantom{0}5.3, \phantom{0}7.6) & \phantom{0}6.2 (\phantom{0}5.7, \phantom{0}7.7) & \phantom{0}3.9 (\phantom{0}3.4, \phantom{0}4.1) & \phantom{0}5.7 (\phantom{0}5.4, \phantom{0}6.1) \\
3 & US & $\delta$ Sco & 691 & 485 & 505 & \phantom{0}9.8 (\phantom{0}8.4, 11.0) & 10.5 (\phantom{0}8.6, 11.0) & \phantom{0}6.4 (\phantom{0}5.8, \phantom{0}7.5) & \phantom{0}7.8 (\phantom{0}7.4, \phantom{0}8.8) \\
4 & US & $\beta$ Sco & 285 & 200 & 206 & \phantom{0}7.6 (\phantom{0}6.9, \phantom{0}8.4) & \phantom{0}8.1 (\phantom{0}7.6, \phantom{0}8.4) & \phantom{0}4.8 (\phantom{0}4.6, \phantom{0}5.3) & \phantom{0}7.0 (\phantom{0}6.7, \phantom{0}7.6) \\
5 & US & $\sigma$ Sco & 544 & 381 & 415 & 10.0 (\phantom{0}9.5, 11.0) & 10.5 (\phantom{0}9.4, 11.0) & \phantom{0}6.0 (\phantom{0}5.8, \phantom{0}6.3) & \phantom{0}8.9 (\phantom{0}8.4, \phantom{0}9.4) \\
6 & US & Antares & 502 & 347 & 357 & 12.7 (11.0, 13.1) & 12.6 (10.3, 12.9) & \phantom{0}6.0 (\phantom{0}5.8, \phantom{0}6.5) & \phantom{0}8.9 (\phantom{0}8.5, \phantom{0}9.5) \\
7 & US & $\rho$ Sco & 240 & 173 & 180 & 13.7 (13.1, 15.0) & 13.7 (13.0, 15.0) & \phantom{0}7.9 (\phantom{0}7.4, \phantom{0}8.3) & 11.5 (10.4, 12.4) \\
8 & US & Scorpio-Body & 373 & 262 & 268 & 14.7 (14.0, 15.5) & 14.0 (13.6, 15.5) & \phantom{0}9.1 (\phantom{0}8.8, \phantom{0}9.5) & 13.2 (12.6, 13.8) \\
9 & US & US-foreground & 276 & 204 & 210 & 19.1 (17.8, 21.5) & 18.4 (17.7, 19.4) & 12.5 (12.1, 13.1) & 17.4 (16.6, 18.2) \\
\hline 10 & UCL & V1062-Sco & 1029 & 760 & 778 & 15.0 (13.6, 15.9) & 15.1 (13.1, 15.9) & 10.0 (\phantom{0}9.8, 10.5) & 13.2 (12.4, 13.7) \\
11 & UCL & $\mu$ Sco & 54 & 43 & 44 & 17.2 (14.8, 18.1) & 15.3 (14.8, 18.0) & \phantom{0}9.6 (\phantom{0}9.0, 10.5) & 13.6 (12.4, 15.8) \\
12 & UCL & Libra-South & 71 & 54 & 57 & 20.0 (17.8, 22.5) & 19.4 (17.8, 22.7) & 12.4 (10.8, 13.3) & 18.9 (17.3, 21.8) \\
13 & UCL & Lupus 1-4 & 226 & 137 & 138 & \phantom{0}6.0 (\phantom{0}5.1, \phantom{0}6.6) & \phantom{0}5.9 (\phantom{0}4.8, \phantom{0}6.5) & \phantom{0}4.3 (\phantom{0}4.1, \phantom{0}5.2) & \phantom{0}6.3 (\phantom{0}5.7, \phantom{0}7.1) \\
14 & UCL & $\eta$ Lup & 769 & 580 & 598 & 15.3 (15.0, 15.9) & 14.8 (14.1, 15.5) & \phantom{0}9.8 (\phantom{0}9.1, 10.1) & 14.4 (13.8, 14.7) \\
15 & UCL & $\phi$ Lup & 1114 & 792 & 825 & 16.9 (16.3, 17.8) & 17.7 (16.4, 17.9) & \phantom{0}9.9 (\phantom{0}9.4, 10.3) & 16.4 (15.9, 17.0) \\
16$^b$ & UCL & Norma-North & 42 & 32 & 33 & 42.1 (33.2, 51.1) & 35.3 (29.2, 51.2) & 38.8 (29.9, 42.9) & 36.0 (28.5, 42.5) \\
17 & UCL & $e$ Lup & 516 & 401 & 413 & 20.9 (20.1, 21.6) & 21.4 (20.8, 22.3) & 11.0 (10.7, 11.5) & 17.9 (16.4, 20.1) \\
18 & UCL & UPK606 & 131 & 97 & 100 & 13.4 (12.7, 14.8) & 13.2 (12.7, 14.7) & \phantom{0}8.5 (\phantom{0}8.0, \phantom{0}8.9) & 12.5 (11.7, 13.2) \\
19 & UCL & $\rho$ Lup & 246 & 182 & 190 & 14.4 (13.5, 14.8) & 14.4 (13.4, 14.8) & \phantom{0}9.4 (\phantom{0}8.9, 10.2) & 13.8 (13.0, 14.1) \\
20 & UCL & $\nu$ Cen & 1737 & 1265 & 1335 & 15.7 (14.8, 16.0) & 15.2 (14.9, 16.1) & \phantom{0}9.5 (\phantom{0}8.8, 12.1) & 14.6 (14.3, 15.1) \\
\hline 21 & LCC & $\sigma$ Cen & 1805 & 1308 & 1342 & 15.5 (15.0, 16.1) & 15.5 (15.1, 15.9) & \phantom{0}9.5 (\phantom{0}8.8, \phantom{0}9.9) & 14.7 (14.4, 15.3) \\
22 & LCC & Acrux & 394 & 276 & 283 & 11.2 (10.2, 12.2) & 10.7 (10.1, 11.5) & \phantom{0}7.3 (\phantom{0}6.9, \phantom{0}7.4) & 10.4 (10.1, 10.8) \\
23 & LCC & Musca-foreground & 95 & 66 & 67 & 10.2 (\phantom{0}9.5, 11.2) & 10.3 (\phantom{0}9.5, 11.8) & \phantom{0}6.9 (\phantom{0}6.7, \phantom{0}7.4) & 10.7 (\phantom{0}9.8, 11.7) \\
24 & LCC & $\epsilon$ Cham & 39 & 23 & 24 & \phantom{0}8.8 (\phantom{0}8.4, \phantom{0}9.4) & \phantom{0}8.8 (\phantom{0}8.4, 10.8) & \phantom{0}5.4 (\phantom{0}5.0, \phantom{0}6.0) & \phantom{0}7.9 (\phantom{0}7.1, \phantom{0}9.6) \\
25 & LCC & $\eta$ Cham & 30 & 20 & 21 & \phantom{0}9.4 (\phantom{0}8.5, 10.8) & \phantom{0}8.8 (\phantom{0}8.6, 10.8) & \phantom{0}6.1 (\phantom{0}5.5, \phantom{0}6.9) & \phantom{0}8.6 (\phantom{0}7.4, 10.5) \\
\hline 26 & Pipe & B59 & 32 & 15 & 15 & \phantom{0}3.4 (\phantom{0}2.5, \phantom{0}6.5) & \phantom{0}6.5 (\phantom{0}3.2, \phantom{0}6.9) & \phantom{0}3.0 (\phantom{0}2.1, \phantom{0}3.1) & \phantom{0}3.1 (\phantom{0}2.6, \phantom{0}4.4) \\
27 & Pipe & Pipe-North & 42 & 31 & 31 & 15.9 (13.8, 17.5) & 16.5 (13.1, 19.9) & 10.1 (\phantom{0}8.4, 10.8) & 13.3 (12.1, 16.7) \\
28 & Pipe & $\theta$ Oph & 98 & 70 & 70 & 15.4 (13.5, 16.2) & 14.1 (13.4, 15.9) & \phantom{0}9.5 (\phantom{0}8.9, 10.5) & 15.6 (14.0, 16.3) \\
\hline 29 & CrA & CrA-Main & 96 & 65 & 69 & \phantom{0}8.5 (\phantom{0}6.1, 10.5) & \phantom{0}9.1 (\phantom{0}8.1, 11.3) & \phantom{0}6.1 (\phantom{0}5.5, \phantom{0}8.5) & \phantom{0}8.6 (\phantom{0}7.4, 10.6) \\
30 & CrA & CrA-North & 351 & 243 & 271 & 11.6 (10.8, 12.1) & 11.2 (10.8, 11.8) & \phantom{0}6.6 (\phantom{0}5.8, \phantom{0}6.9) & \phantom{0}9.9 (\phantom{0}9.2, 10.9) \\
31 & CrA & Scorpio-Sting & 132 & 92 & 94 & 14.5 (13.9, 15.1) & 13.9 (12.6, 14.5) & \phantom{0}8.5 (\phantom{0}8.1, \phantom{0}9.0) & 11.3 (10.6, 12.3) \\
\hline 32 & Cham & Centaurus-Far & 99 & 66 & 68 & \phantom{0}8.5 (\phantom{0}7.2, \phantom{0}9.6) & \phantom{0}8.0 (\phantom{0}7.2, \phantom{0}9.6) & \phantom{0}5.9 (\phantom{0}5.6, \phantom{0}7.1) & \phantom{0}9.0 (\phantom{0}8.1, 11.1) \\
33 & Cham & Chamaeleon-1 & 192 & 107 & 114 & \phantom{0}3.8 (\phantom{0}2.9, \phantom{0}5.7) & \phantom{0}3.5 (\phantom{0}2.7, \phantom{0}4.3) & \phantom{0}3.0 (\phantom{0}2.6, \phantom{0}3.8) & \phantom{0}3.8 (\phantom{0}3.3, \phantom{0}4.3) \\
34 & Cham & Chamaeleon-2 & 54 & 24 & 24 & \phantom{0}2.8 (\phantom{0}1.7, \phantom{0}3.5) & \phantom{0}2.1 (\phantom{0}1.7, \phantom{0}2.9) & \phantom{0}2.8 (\phantom{0}2.4, \phantom{0}3.1) & \phantom{0}3.1 (\phantom{0}2.5, \phantom{0}4.1) \\
\hline 35 & NE & L134/L183 & 24 & 11 & 11 & \phantom{0}9.6 (\phantom{0}7.4, 11.3) & \phantom{0}9.7 (\phantom{0}7.2, 13.6) & \phantom{0}5.7 (\phantom{0}5.1, \phantom{0}6.6) & \phantom{0}9.4 (\phantom{0}7.9, 11.7) \\
36$^b$ & NE & Oph-Southeast & 61 & 49 & 52 & \phantom{0}9.2 (\phantom{0}7.5, 12.5) & \phantom{0}8.0 (\phantom{0}7.2, \phantom{0}9.2) & \phantom{0}9.5 (\phantom{0}8.5, \phantom{0}9.7) & \phantom{0}9.5 (\phantom{0}9.0, 11.3) \\
37$^b$ & NE & Oph-NorthFar & 28 & 20 & 22 & 19.1 (14.5, 25.7) & 19.4 (14.9, 24.2) & \phantom{0}9.8 (\phantom{0}8.5, 14.9) & 17.5 (14.1, 21.0) \\

\hline
\end{tabular}
} 
\renewcommand{\arraystretch}{1}
\label{tab:ages}
\tablefoot{The ages are given for the four isochronal fitting results using the PARSEC and BHAC15 models for both the BPRP and GRP CMDs. The lower and upper age limits are given in parenthesis, determined from the 1$\sigma$ highest density interval from the marginalized posterior PDF. 
The differently determined ages are compared in Fig.~\ref{fig:age_splom}.
The clusters are grouped into traditional subregions for orientation, motivated by the original Blaauw borders, as described in \citetalias{Ratzenboeck2022}, while these separations should generally not be treated as physically meaningful entities.
\tablefoottext{a}{The numbers of sources are as follows: All = the number of all stellar member candidates per \texttt{SigMA} cluster. BPRP = after applying the photometric quality criteria for the BPRP CMD. GRP = after applying the photometric quality criteria for the GRP CMD.}
\tablefoottext{b}{The three clusters Norma-North, Oph-SE, and Oph-NF are excluded from the analysis of the star formation history in Sco-Cen since they are likely unrelated (see text).}
}
\end{small}
\end{center}
\end{table*}

In this paper, we determine robust isochronal ages for the 37 \texttt{SigMA} clusters in Sco-Cen, which contain 13,103 candidate Sco-Cen members. A detailed description of the process used to select the cluster sample can be found in \citetalias{Ratzenboeck2022} (\texttt{SigMA} algorithm), and an overview of the clusters is presented in Table~\ref{tab:ages}. The clusters are assigned to traditional sub-regions within Sco-Cen, including the classical \citet{Blaauw1946} definition (see details in \citetalias{Ratzenboeck2022} and Table~\ref{tab:ages}). However, we point out that these borders were drawn initially in 2D on the plane of the sky and should not be seen as physically meaningful entities. They should instead be used for orientation and comparisons with previous works. 

To determine the ages of the Sco-Cen clusters, we utilized two different evolutionary model families: PARSEC v1.2S~\citep{Bressan2012, Chen2015parsec, Marigo2017parsec} (hereafter PARSEC) and \citet{Baraffe2015} (hereafter BHAC15). These models are outlined in more detail in Appendix~\ref{apx:method}. We also use two different color-absolute magnitude diagrams (CMDs) based on \textit{Gaia} photometric systems: ($M_G$ versus $G_\mathrm{BP} - G_\mathrm{RP}$) and ($M_G$ versus $G - G_\mathrm{RP}$), which are abbreviated as BPRP and GRP, respectively. The absolute magnitude $M_G$ was calculated using the distance modulus using the inverse of the parallax as distance, which is reasonable for sources with Sco-Cen distances and low uncertainties, as discussed in \citetalias{Ratzenboeck2022}.

We applied a set of photometric quality criteria to the \textit{Gaia} DR3 photometry to achieve more reliable isochronal age fitting. The influence of photometric uncertainties is highlighted in Fig.~\ref{fig:hrd-all}. We use the corrected flux excess factor $C^*$ as described in \citet{Riello2021}. As noted in \citet{Evans2018}, large values of the flux excess factor are the result of issues in the $G_\mathrm{BP}$ or $G_\mathrm{RP}$ photometry. Additionally, we cut photometric flux errors $G_\mathrm{err}$, $G_\mathrm{BP,err}$, and $G_\mathrm{RP,err}$, as well as RUWE, which (preferentially) removes unresolved binaries in the sample \citep{Lindegren2018, Lindegren2021}. 
The parameters used are summarized in Eq.\,(\ref{eq:errs}) and the quality criteria in Eq.\,(\ref{eq:phot-cut}).

\begin{equation}
    \label{eq:errs}
    \begin{aligned}
        & G_\mathrm{err} = 1.0857/\mathtt{phot\_g\_mean\_flux\_over\_error} \\
        & G_\mathrm{RP,err} = 1.0857/\mathtt{phot\_rp\_mean\_flux\_over\_error}  \\
        & G_\mathrm{BP,err} = 1.0857/\mathtt{phot\_bp\_mean\_flux\_over\_error} \\
        & \text{RUWE} = \text{re-normalised unit weight error} \\ 
        & C = \mathtt{phot\_bp\_rp\_excess\_factor} =  (I_\mathrm{BP} + I_\mathrm{RP})/I_G \\
        & C^* = \text{corrected} \,\, C \\
        & \sigma_{C^*} = 0.0059898 + 8.817481 \times 10^{-12} \cdot G^{7.618399} 
    \end{aligned}
\end{equation} 

\noindent 
Our sample is restricted to sources that satisfy the following photometric quality criteria:
\begin{equation}
    \label{eq:phot-cut}
    \begin{aligned}
        & G_\mathrm{err} < 0.007 \,\text{mag} \\
        & G_\mathrm{RP,err} < 0.03 \,\text{mag} \\
        & G_\mathrm{BP,err} < 0.15 \,\text{mag} \\
        & \mathrm{RUWE} < 1.4 \\
        & (C^* < 5 \, \sigma_{C^*} \,\,\, \text{AND} \,\,\,  G>5\,\text{mag}) \,\,\,  \text{OR} \,\,\,  G \leq 5\,\text{mag}
    \end{aligned}
\end{equation}

We use these criteria for the BPRP CMD, while we exclude the $G_\mathrm{BP,err}$ condition when using the GRP CMD. We visually confirm that these quality criteria reduce the scatter around isochrones significantly, especially in the low-mass regime (see Fig.~\ref{fig:hrd-all}).
We further constrain the absolute magnitude range of the sample, using different cuts for the two model families. 
\begin{equation}
    \label{eq:mag-cut-parsec}
    M_G < 10\,\text{mag} \,\, \text{(when using PARSEC models)}
\end{equation}
\begin{equation}
    \label{eq:mag-cut-bhac}
    M_G < 12\,\text{mag} \,\, \text{(when using BHAC15 models)} 
\end{equation}
These cuts are motivated by our observation of areas in the CMD with varying degrees of overlap between data and model isochrones (see Figs.~\ref{fig:BPRP-HRDs-1}--\ref{fig:GRP-HRDs-2}). Furthermore, the PARSEC and BHAC15 isochrones start to disagree with each other towards the faint end, while the BHAC15 models tend to agree better with the data fainter than $M_G \gtrsim 10$\,mag compared to PARSEC models. Therefore, we cut at $M_G<10$\,mag to reduce systematic age shifts determined with PARSEC. This choice is also supported by the mass coverage of the PARSEC isochrones, which are cut off at 0.09\,M$_\odot$, while BHAC15 includes low-mass objects down to 0.01\,M$_\odot$. 
The magnitude limit for the BHAC15 models is motivated by the larger uncertainties of the observations in the faint regime, which increases the scatter in the CMD below 12\,mag (see Fig.~\ref{fig:hrd-all}).

Applying the photometric quality criteria from Eq.\,(\ref{eq:phot-cut}) to the Sco-Cen sample of 13,103 stars, there are 9,249 sources (71\%) remaining in the BPRP CMD, and 9,593 (73\%) in the GRP CMD (see Fig.~\ref{fig:hrd-all}). When applying the additional magnitude limits in Eq.\,(\ref{eq:mag-cut-parsec}), there are 5,257 ($\sim$40\%) sources left in both CMDs for PARSEC-BPRP and PARSEC-GRP.    
For BHAC15-BPRP and BHAC15-GRP there are 8,514 and 8,521 sources left, respectively ($\sim$65\% each), after applying the magnitude cut from Eq.\,(\ref{eq:mag-cut-bhac}). The BHAC15 isochrones do not cover the upper main-sequence, since the models stop around $M_G\sim4$\,mag (at 1.4\,M$_\odot$). However, this is not an issue for the age fitting method, since the method uses only sources until the maximum brightness of each fitted isochrone. This also reduces the number of sources used for fitting to BHAC15 isochrones, depending on the maximum $M_G$.

\section{Method} \label{sec:method}

Determining an accurate age for each cluster is critical for our goal of distinguishing small age differences between clusters and creating a high-resolution Sco-Cen age map. Our isochrone fitting procedure is summarized here, with a comprehensive description in Appendix~\ref{apx:method}. 

Rather than simply minimizing the least sum of squares between data points and isochronal curves, we aim to account for observational trends such as unresolved binaries and extinction. We assume a simple model in which data are generated along isochrones with noise contributions drawn independently from skewed Cauchy distributions with zero means to model non-symmetric noise sources. The influence of reddening and unresolved binaries on the displacement of sources in the CMD is different for each cluster. Instead of fixing the skewness and scale parameters of the skewed Cauchy distribution, we let them be free parameters of the model, which are obtained during Bayesian inference alongside astrophysical parameters such as cluster age and dust extinction. This age fitting technique is explained in detail in Appendix~\ref{apx:method}.

\section{Results} \label{sec:resutls}

We present robust isochronal ages for the 37 \texttt{SigMA} clusters in Sco-Cen as selected in \citetalias{Ratzenboeck2022}. We provide four different age estimates for the clusters, determined with PARSEC-BPRP, PARSEC-GRP, BHAC15-BPRP, and BHAC15-GRP, as outlined in Sect.~\ref{sec:data} and Appendix~\ref{apx:method}.  The ages determined in this fashion are listed in Table\,\ref{tab:ages}.
In Appendix~\ref{apx:kerr}, we compare our ages in more detail to the cluster sample by \citet{Kerr2021}, who found a similar substructure in Sco-Cen, while with lower numbers statistics. The clusters in \citet{Kerr2021} appear to be systematically older compared to our age estimates, while we do not (yet) have a clear explanation for this behavior.

In Appendix~\ref{aux:fig}, we provide the CMDs showing the best-fitting isochrone models for each cluster and the four different age-fitting results. Unless otherwise noted, we adopt the PARSEC-BPRP ages in our analysis. These isochrones seem robust within the errors compared to the PARSEC-GRP and BHAC15-GRP ages. Furthermore, the combination of PARSEC-BPRP is often used in the literature, which facilitates comparison to previous work \citep[e.g.,][]{Bossini2019, Dias2019, Cantat-Gaudin2020b, Kerr2021}. 
 
We exclude three clusters from the original \texttt{SigMA} sample of 37 clusters, namely Norma-North (Norma-N), Oph-Southeast (Oph-SE), and Oph-NorthFar (Oph-NF), leaving 34 clusters for further discussion in Sect.\,\ref{sec:history}. Norma-N is excluded, as it is substantially older ($\sim$40\,Myr) than the nominal upper age limit of Sco-Cen, which is about 20\,Myr. Moreover, Norma-N, Oph-SE, and Oph-NF appear to be kinematically unrelated as suggested by the trace-backs of the cluster orbits, which is discussed in more detail in a follow-up paper (Gro{\ss}schedl et al. in prep.).

Figure~\ref{fig:ScoCen3D-ages} presents the spatial distribution of the surfaces enveloping the 34 clusters, color-coded by cluster ages. This figure can be investigated as an interactive 3D plot\footnote{See the link to the \href{https://homepage.univie.ac.at/sebastian.ratzenboeck/wp-content/uploads/2023/05/scocen_age.html}{interactive 3D version}.}, which best illustrates the spatial arrangement of Sco-Cen's clusters. It provides important insight into how the complex was assembled. At first glance, it is easy to see that the center of the association contains the oldest clusters, while the youngest clusters, appearing in blue, tend to be at the outskirts of the association. The existence of patterns of age gradients across contiguously located clusters is also clear. 

The age gradients are particularly clear towards two quasi-vertical, approximately 100-pc long chains of clusters. These contiguous cluster chains connect the older clusters at the (older) center of the association with the younger clusters at the (younger) outskirts. The first one comprises the traditional Blaauw's LCC (from here on, LCC chain), while the second connects CrA to the center of the association (from here on, CrA chain). They are remarkable structures for being a contiguous series of clusters following a coherent age gradient. The LCC chain clusters comprise, from old to young and north to south: $\sigma$\,Cen, Acrux, Musca-foreground, $\epsilon$\,Cham, and $\eta$\,Cham. 
The member clusters toward the CrA chain are less clear and are tentatively arranged as follows, from old to young: $\phi$\,Lup, $\eta$\,Lup, (V1062\,Sco, $\mu$\,Sco)\footnote{V1062\,Sco and $\mu$\,Sco are located slightly off from the chain, towards the back of Sco-Cen. Future work is needed to understand the true connections between clusters.}, Sco-Body, Sco-Sting, CrA-North, and CrA-Main.  

Toward the Galactic northeast, the main body of Sco-Cen is connected to USco in a similar manner to the cluster chains (see Fig.~\ref{fig:3Dtimeline}), forming a third chain of clusters, albeit more complex. Recent attempts to reconstruct the star formation history of the USco region also reveal a more complex substructure compared to the LCC or CrA chains \citep[e.g,][]{Squicciarini2021, MiretRoig2022b, BricenoMorales2023}. This could possibly be explained by a different original gas distribution and by the number and distribution of massive stars located within USco itself or the older clusters in Sco-Cen \citep[e.g.,][]{Diehl2010, Robitaille2018, Krause2018, Neuhaeuser2020, Forbes2021}. The origin of USco warrants a dedicated analysis beyond the scope of this paper, but it seems clear from Fig.~\ref{fig:3Dtimeline} that USco is the third chain of clusters. These three chains started to form about 10\,Myr ago, and are likely induced or enhanced by the feedback generated by the massive clusters formed about 15\,Myr ago.

\begin{figure}[!t]
    \centering
    \includegraphics[width=0.99\columnwidth]{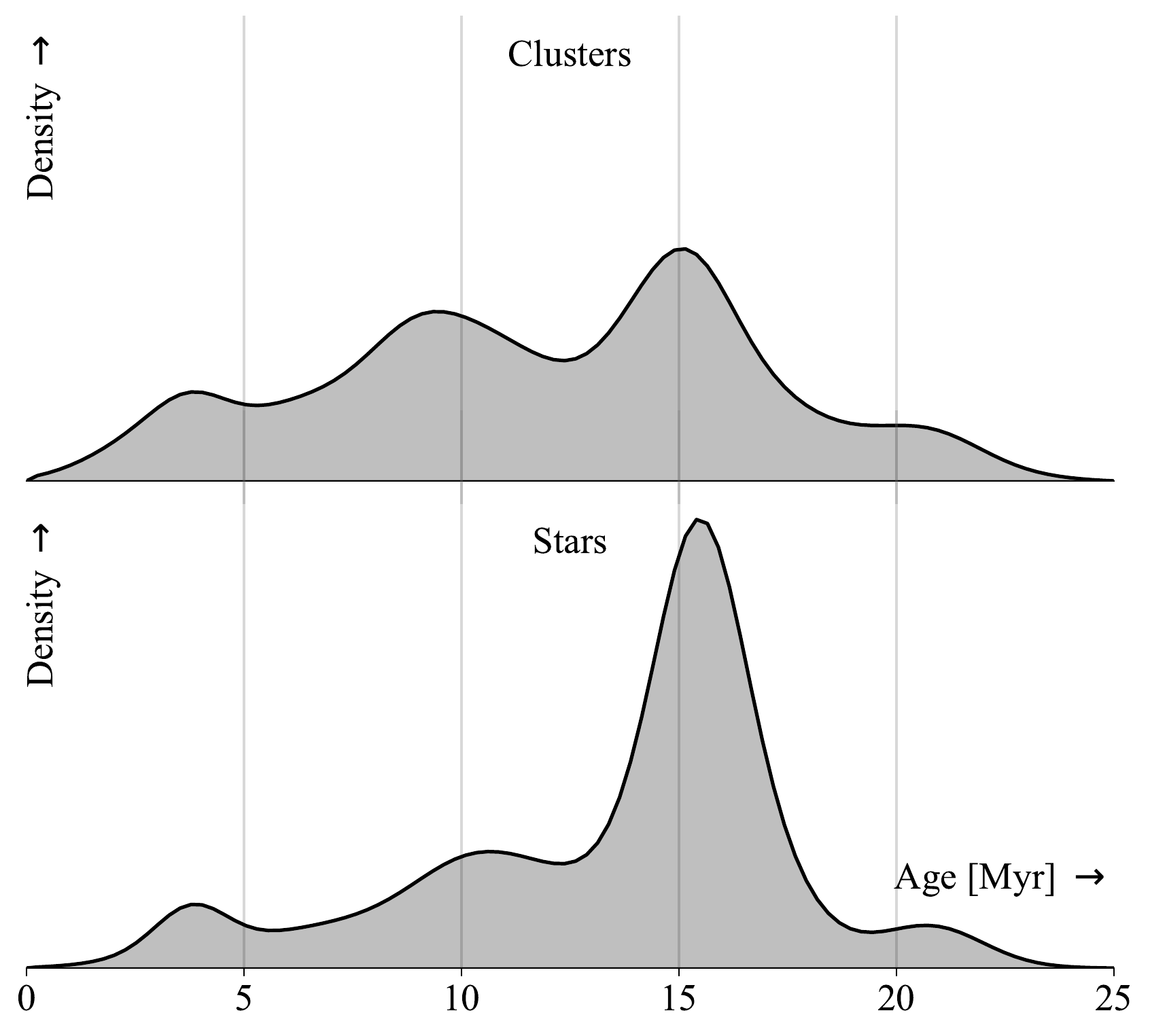}
    \caption{Star formation history of Sco-Cen. Top: The age distribution of the 34 clusters in Sco-Cen traces four main star formation events in the history of the association. Bottom: The stellar age distribution in Sco-Cen shows a similar pattern. To study the age distribution of clusters and stars, we used a kernel density estimate with an adaptive bandwidth corresponding to the age uncertainty of each sample. The formation of stars and clusters is correlated, but not linearly, meaning that, when compared to the average in the association, more stars were formed per cluster during the peak of star formation rate, about 15 Myr ago.} 
    \label{fig:age-distribution}
\end{figure}

\begin{figure*}[!t]
    \includegraphics[width=2.05\columnwidth]{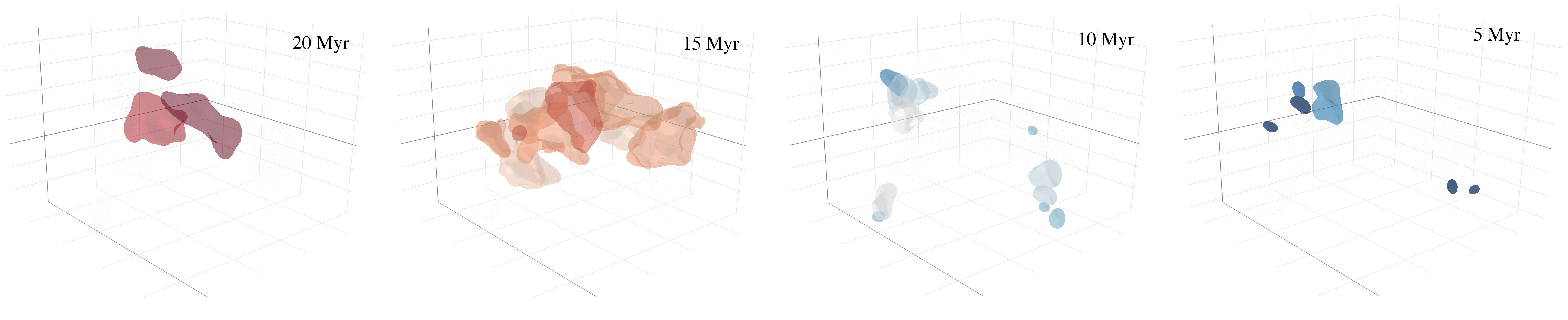}
    \caption{Star formation progression in Sco-Cen, shown with the same orientation in XYZ and color-scaling as in Fig.\,\ref{fig:ScoCen3D-ages}. By separating the clusters in age bins, following the peaks in Fig.~\ref{fig:age-distribution} and Table~\ref{tab:bursts}, one can appreciate a consistent inside-out progression of star formation from older to younger clusters. For more details, see the link to the \href{https://homepage.univie.ac.at/sebastian.ratzenboeck/wp-content/uploads/2023/05/scocen_age.html}{interactive 3D version}.} 
    \label{fig:3Dtimeline}
\end{figure*}

\begin{table*}[!t] 
\begin{center}
\begin{small}
\caption{Overview of the star formation modes/bursts as determined from the age distribution of Sco-Cen clusters (Fig.~\ref{fig:age-distribution}).\vspace{-2.5mm}}
\renewcommand{\arraystretch}{1.3}
\resizebox{1.8\columnwidth}{!}{
\begin{tabular}{llp{45mm}p{80mm}} 

\hline \hline

\multicolumn{1}{c}{Notation} &
\multicolumn{1}{c}{Age (Myr)} &  
\multicolumn{1}{c}{Description} &
\multicolumn{1}{c}{Clusters\tablefootmark{a}} \\
\cmidrule(lr){1-4}

20\,Myr peak & $\sim$20--22 & Initial onset of star formation in Sco-Cen & $e$\,Lup, Libra-South, US-foreground \\

15\,Myr peak & (14.4, 15.9) & Maximum of star and cluster formation rate & $\nu$\,Cen, $\sigma$\,Cen, $\rho$\,Lup, UPK606, V1062-Sco, $\mu$\,Sco, $\phi$\,Lup, $\eta$\,Lup, Pipe-N, $\theta$\,Oph, Sco-Body, Sco-Sting, $\rho$\,Sco \\

10\,Myr peak & (8.5, 10.2) & Formation of the major clusters in USco and the chains of LCC and CrA & Antares, $\sigma$\,Sco, $\delta$\,Sco, $\beta$\,Sco, L134/L183, Acrux, Musca-fg, $\epsilon$\,Cham, $\eta$\,Cham, Cen-Far, CrA-North, CrA-Main\tablefootmark{b}  \\

5\,Myr peak & $\sim$3--5 & Recent star formation, still near or associated with gas & Lupus\,1-4, $\nu$\,Sco, $\rho$\,Oph, B59, Cham-1, Cham-2 \\


\hline
\end{tabular}
} 
\renewcommand{\arraystretch}{1}
\label{tab:bursts}
\tablefoot{There are two main modes of star formation, including the second and third peak at 15\,Myr and 10\,Myr ago, which are confirmed as significant by the excess of mass test. We find two smaller peaks at 20\,Myr and 5\,Myr ago, for which we give the time-frames approximately.
\tablefoottext{a}{We tentatively assign each of the 34 \texttt{SigMA} clusters to one of the peaks, as separately displayed in Fig.\,\ref{fig:3Dtimeline}.}
\tablefoottext{b}{CrA-Main is likely closely related to the younger Coronet cluster, which is largely embedded in the head of the CrA molecular cloud.}
}
\end{small}
\end{center}
\end{table*}

Figure~\ref{fig:age-distribution} presents the age distribution within the Sco-Cen complex. In the upper panel, we show the distribution of cluster ages using the selected 34 \texttt{SigMA} clusters. In the lower panel, we display the age distribution for the stellar members by assigning each star the age of the parent cluster. 
We employ a kernel density estimate (KDE) technique to examine the age distribution of clusters and stars, focusing on uncovering the modality structure of the probability density function (PDF). To account for age uncertainty, we implemented an adaptive KDE in which the kernels' size reflects the determined age variance of the samples (see age uncertainties in Table~\ref{tab:ages}). We have standardized the area under the curves to represent the number of clusters and stars for the respective PDFs. 

The distribution of cluster ages in Fig.~\ref{fig:age-distribution} exhibits a significant multi-modality. When applying the calibrated dip test~\citep{Hartigan:1985, Cheng:1998} to the cluster ages, we find a strong indication ($p$\,$<$\,0.05) for multiple modes in the data. To locate the modes, we apply the excess of mass test~\citep{Muller:1991, Cheng:1998}, which finds two modes in the following age ranges: (8.5, 10.2), (14.4, 15.9) Myr (see Fig.~\ref{fig:age-distribution}, upper panel). Together they make up the two main modes of star formation in the evolution of Sco-Cen, in which over 60\% of stars have been formed. The adaptive KDE in Fig.~\ref{fig:age-distribution} reveals a small third and fourth peak at around 4\,Myr and 21\,Myr, respectively. The four peaks in star formation rate are summarized in Table~\ref{tab:bursts}.

\section{Discussion} \label{sec:discussion}

We first examine the spatial-temporal patterns in Sco-Cen revealed by isochrone fitting for the individual clusters (Sect.~\ref{sec:history}) and consider potential explanations for the observed age patterns (Sect.~\ref{sec:octopus}). In Sect.~\ref{sec:fang}, we provide a solution to the Upper-Sco age controversy that has perplexed the literature for decades, clarified by \textit{Gaia} data.

\subsection{Spatial-temporal patterns in Sco-Cen} \label{sec:history}

The spatial-temporal arrangement of the clusters in Fig.~\ref{fig:ScoCen3D-ages} indicates that star formation in Sco-Cen did not proceed chaotically. Figure~\ref{fig:3Dtimeline} shows groups of clusters associated with the star formation rate peaks from Fig.~\ref{fig:age-distribution} and Table\,\ref{tab:bursts}. This figure shows a clear relationship between age and position of Sco-Cen clusters, with the older clusters (age $>$\,15\,Myr) towards the association’s center and the younger cluster towards the outskirts of the association. This spatial arrangement implies an inside-out star formation scenario for Sco-Cen and evokes a feedback-driven scenario reminiscent of the canonical triggering scenario in \citet{Elmegreen1977}.

Although we could speculate that the oldest clusters in the association's center ($e$\,Lup and $\phi$\,Lup) provided the supernovae (SNe) for the initial trigger, the masses of these clusters suggest that they might have produced only a few SNe, assuming a normal initial mass function (IMF). A dedicated model, like the forward modeling done for the Ophiuchus region by \citet{Forbes2021}, is warranted to test this suggestion. Perhaps more impressive is the likely number of SNe provided by the $15$\,Myr star formation burst, which could be in the tens of SNe when integrating over the massive clusters formed during this high star formation rate period. These SNe, together with stellar winds, ionizing radiation, and mass loss events, would have injected substantial energy and momentum into the primeval gas in the Sco-Cen region, likely pushing part of it to collapse. 

\begin{figure*}[!t]
    \centering
    \includegraphics[width=2.05\columnwidth]{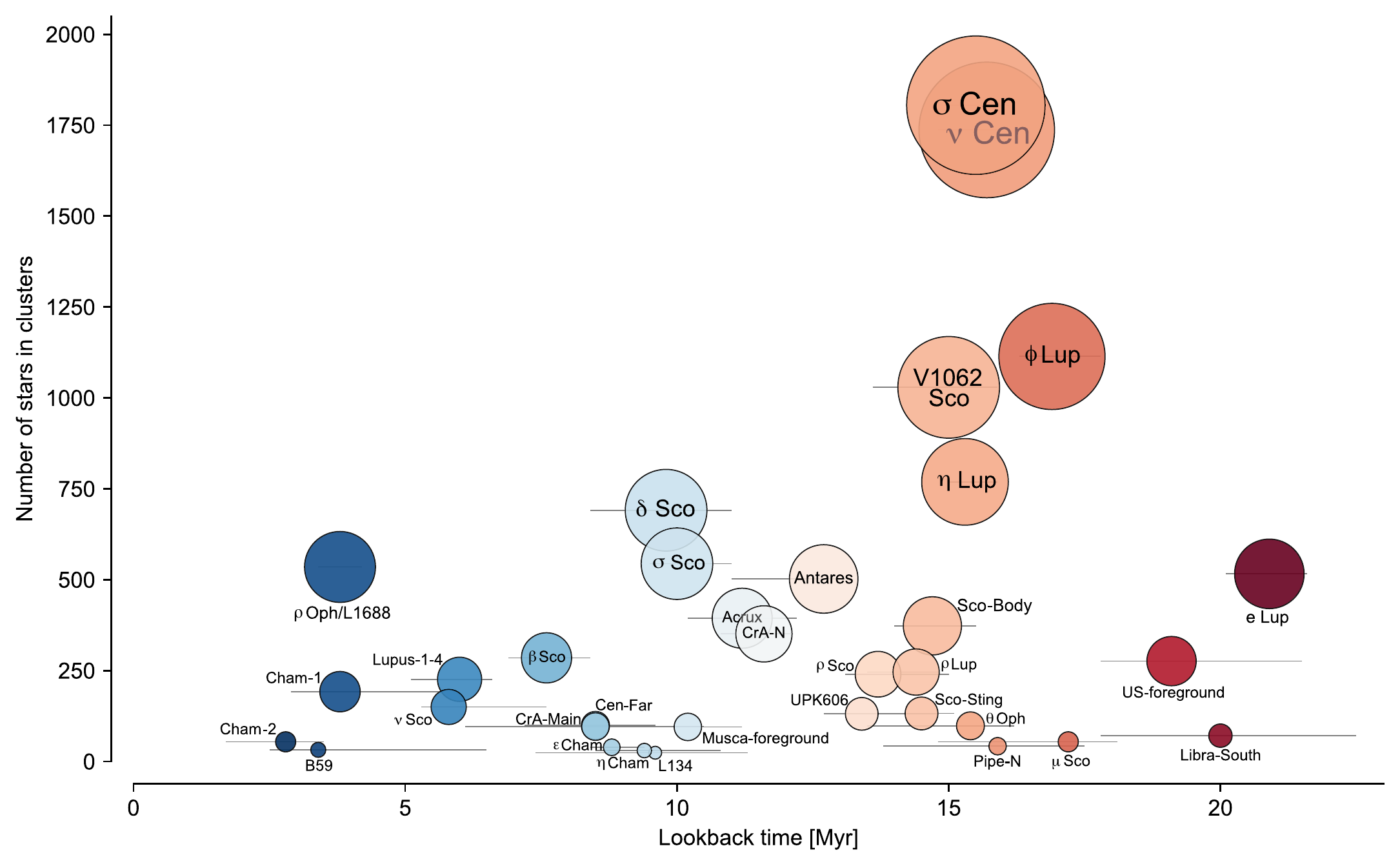}
    \caption{Timeline of the formation of clusters in Sco-Cen, color-coded by age (same as in Fig.\,\ref{fig:ScoCen3D-ages}) and scaled by cluster size. The horizontal lines represent age uncertainties. While Sco-Cen has been forming small clusters ($\leq$ 250 stars) continuously since its formation 20--25\,Myr ago, its large clusters ($\gtrsim$ 1,000 stars) were all formed during the peak of the star formation rate, around 15\,Myr ago. Regarding their absolute numbers, most of the stars and clusters we can observe today in the Sco-Cen region were formed around 15 Myr ago. Since then, star formation has been declining. The peaks of star and cluster formation rate seem to be periodically distributed (every $\sim$\,5\,Myr, see also Fig.~\ref{fig:age-distribution}).} 
    \label{fig:timeline}
\end{figure*}

Figure~\ref{fig:timeline} displays the ages as a function of the number of stars per cluster, color-coded by age and scaled by the number of stars per cluster. This visualization facilitates linking individual clusters to peaks in the star formation history of the region. Similarly to Fig.~\ref{fig:age-distribution}, we can see that a peak at around 15\,Myr ago contributed the largest numbers of stars and the most massive clusters to the association. Since then, star formation has been declining, although periods of increased star formation rate seem to appear about every $\sim$\,5\,Myr (Fig.~\ref{fig:age-distribution}). Small clusters have formed throughout the entire history of Sco-Cen.

By using the distribution and ages of the \texttt{SigMA} clusters, star formation in Sco-Cen can be described as follows: The first Sco-Cen stars were formed about 20--25\,Myr ago in the primordial Sco-Cen giant molecular cloud. At around 15\,Myr ago, there was a burst of star and cluster formation where most stars in the association formed. This burst aligns with a scenario presented in \citet{Zucker2022}, which suggests that the Local Bubble was triggered by massive stellar feedback originating from Sco-Cen. \citet{Zucker2022} suggest that the first SNe that powered the Local Bubble happened about 14\,Myr ago, which is roughly compatible with feedback from the oldest stellar populations in Sco-Cen. Considering that the most massive stars require a few million years to explode as SNe, the likely first Sco-Cen SNe originated in the oldest clusters presented in this work, particularly $e$\,Lup and $\phi$\,Lup. The feedback of possible SNe at that time may have triggered the Sco-Cen 15\,Myr burst, which created the most massive clusters in the association. Subsequently, SNe originating from the 15\,Myr clusters likely continue feeding the Local Bubble's expansion, with the last SNe taking place about 2\,Myr ago \citep{Fuchs2006, Breitschwerdt2016, Feige2017, Krause2018, Neuhaeuser2020}. 

While we propose that feedback played a crucial role in the formation of Sco-Cen, it is hard to quantify how crucial it was. At this point, we have no evidence that the formation of the first stars in Sco-Cen was induced by feedback or any external factor to a collapsing molecular cloud. The same can be argued for the origin of the 15\,Myr burst. While tempting to invoke $e$\,Lup or $\phi$\,Lup as the potential progenitors of this burst, this is tentative and modeling is required to make a stronger statement. Still, what is clear now is that star formation since the peak of star formation rate about 15\,Myr ago formed coherent patterns that are best explained as the direct product of feedback. Clusters younger than 10\,Myr are arranged in quasi-linear radial structures with coherent age gradients from old in the center to younger in the outskirts of the association. The observed ``chains of clusters'' (the LCC, CrA, and the USco chain) are clear examples. 

\subsection{The ``Octopus'' model} \label{sec:octopus}

In their study, \citet{Krause2018} proposed a ``Surround and Squash'' scenario for the formation history of Sco-Cen, suggesting that the region formed from a long connected cloud, shaped as an elongated sheet. They argued that superbubbles continuously broke out of this sheet to surround and squash the denser parts of the cloud (while small cloud fractions initially survive in-between), inducing further star formation and creating several shells and superbubbles. The authors also confirmed the existence of a large super-shell around the entire OB association and a nested filamentary super-shell. \citet{Krause2018} suggested that the first SNe occurred at the center of the primordial Sco-Cen cloud, which agrees with our results. They proposed that possible SNe originating from the oldest clusters compressed the surrounding gas, which could have caused the 15\,Myr burst.

The ``Surround and Squash'' model was designed to explain the age ranking of the subgroups, with UCL being the oldest and USco the youngest.  The main assumption of this model is that the Sco-Cen population is described by the three main Blaauw subgroups, which we now know is an oversimplified description of the association. Our results instead point to an ``Octopus'' model, where most stars and clusters are formed at the head of the octopus, and several arms extend radially outward, containing the younger stars and clusters in the association. There is no obvious need for a ``Surround and Squash'' scenario to explain the observations. Although we do not have sufficient evidence to state the likely role of feedback in the formation of the head, the formation of the arms is very likely a product of the feedback from the massive stellar population in the head. 

In conclusion, the star formation patterns described in this work (combined with earlier evidence for massive stellar feedback) suggest an important role for feedback-driven star formation in a manner similar to the classical sequential star formation scenario of \citet{Elmegreen1977}. The observed octopus-like inside-out formation of Sco-Cen provides a simpler explanation compared to the ``Surround and Squash'' model, while both are feedback driven.  
Our results for the Sco-Cen OB association suggest a generally significant role for feedback in the formation and evolution of OB associations. 
Similar evidence can be found in the Orion OB1 association, where also an important role for massive stellar feedback has been found \citep[e.g.,][]{Brown1994, Brown1995, Ochsendorf2015, Grossschedl2021, Swiggum2021, Foley2022arXiv}, as well as in Vela \citep[e.g.,][]{Cantat-Gaudin2019a, Armstrong2022}, Cygnus \citep[e.g.,][]{Quintana2021, Quintana2022}, or Cepheus \citep[e.g.,][]{Kun1987, Szilagyi2023}.

\subsection{Upper-Sco age controversy} \label{sec:fang}
  
There has been a long discussion in the literature about the age of the Upper-Sco Association (USco). Due to its young age, richness, and proximity to Earth, USco is a unique laboratory for early stellar evolution and planet formation studies. Therefore, getting the correct age for USco (or, better said, the ensemble of different coeval clusters that were previously taken as the single population USco) is critical for multiple research fields across astronomical scales. 

Traditionally, the stellar populations toward USco were often sub-structured into two parts, $\rho$\,Oph (partially embedded young stellar objects in the Ophiuchus cloud), and USco.
The age determinations for USco in the literature from the last decades (pre-\textit{Gaia}) fall broadly around two estimates: 5\,Myr and 10--12\,Myr. \citet{de_Geus1989} used the massive stars in USco to determine an age of about 5 Myr, an estimate confirmed in \citet{Preibisch2002}, using the full stellar mass spectrum of USco. \citet{Pecaut2012} on the other hand determined an age of about 10--12\,Myr using intermediate to high-mass stars. 
\citet{Sullivan2021} find an age gradient in USco, suggesting that the observed mass-dependent age gradient can be explained by a population of undetected binary stars. They argue their result supports the previously suggested 10\,Myr age for USco, with a small intrinsic age spread. 

Since the release of \textit{Gaia} data, several updated cluster catalogs have been published for the USco association, including cluster samples by \citet{Squicciarini2021}, \citet{Kerr2021}, \citet{MiretRoig2022b}, \citet{BricenoMorales2023}, and the \texttt{SigMA} sample from \citetalias{Ratzenboeck2022}. The new view of USco reveals multiple clusters projected on the same region of the sky and the previous ages of 5 or 10\,Myr for USco are now superseded. The ages for the overlapping clusters along the same line-of-sight range from 3 to 19\,Myr (see Table~\ref{tab:ages}).

\begin{figure}[!t]
        \includegraphics[width=0.9\columnwidth]{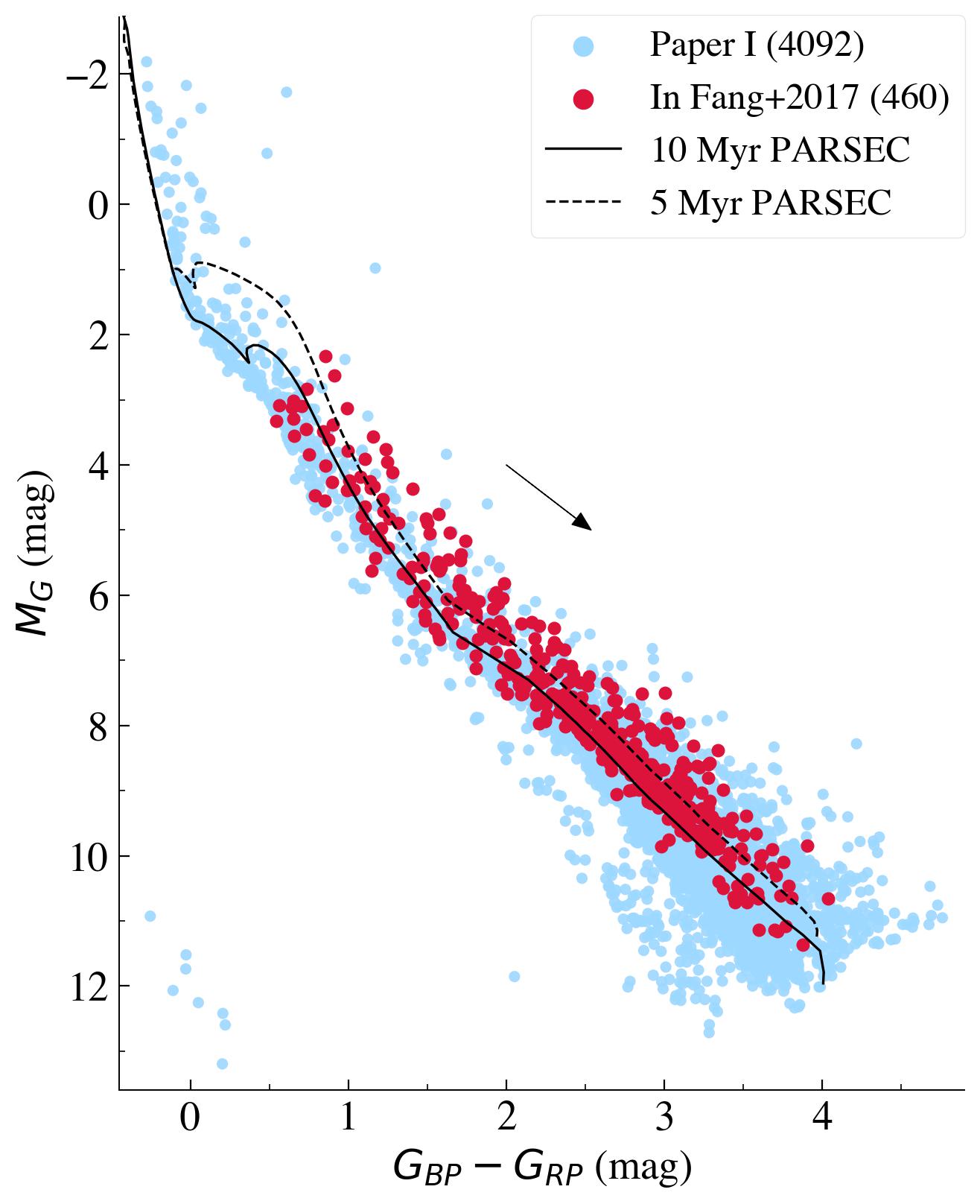}
        \caption{\textit{Gaia} BPRP CMD displaying members of the 12 stellar clusters that have matches with USco members from \citetalias{Fang2017}. The blue dots are all sources in the 12 \texttt{SigMA} clusters that pass additional photometric quality criteria (see text). The red dots indicate sources that are both in \texttt{SigMA} and \citetalias{Fang2017}. Two PARSEC isochrones are shown for 5\,Myr (solid) and 10\,Myr (dashed), marking the earlier assumed nominal ages of USco. The arrow shows an extinction vector with a length of $A_G=1$\,mag. See Fig.~\ref{fig:12-hrds-fang17} for an overview of the separate CMDs of the individual 12 clusters.}
        \label{fig:12-hrds-fang17-all}
    \end{figure}

\citet{Fang2017} (hereafter F17) discuss a sample of stars in the USco region compiled from several sources in the literature \citep[][]{Preibisch1999, Ardila2000, Slesnick2006, Preibisch2002, LuhmanMamajek2012, Rizzuto2015, Pecaut2016}. The sample includes known stellar parameters, such as spectral types, temperatures, and stellar luminosities, allowing for age analysis in the Hertzsprung–Russell diagram (HRD). However, for their discussion, \citetalias{Fang2017} assume that the stars belong to a single population. Treating this sample as a single population generates a large spread in the HRD (see their Fig.~5 and our CMD in Fig.~\ref{fig:12-hrds-fang17-all}), making it impossible to reliably determine the age. For example, isochrones from about 3 to 15\,Myr all provide good fits to the data, depending on the spectral type.

The fact that the \texttt{SigMA} clusters\footnote{The \texttt{SigMA} algorithm does not use any age or color information, only the phase-space density.} and the mentioned other recent clustering studies using \textit{Gaia} data show narrow CMD sequences (hence, better-constrained ages) highlights the high value of the high precision astrometry of the \textit{Gaia} satellite.   
We can now revisit the \citetalias{Fang2017} sample cross-matching it with the \texttt{SigMA} clusters\footnote{Using the clustered substructure from \citet{Squicciarini2021}, \citet{Kerr2021}, \citet{MiretRoig2022b} or \citet{BricenoMorales2023} would deliver similar results.} to investigate the USco age controversy. We find that there are about 500 cross-matches of \citetalias{Fang2017} with \texttt{SigMA}, which are contained in 12 of the \texttt{SigMA} clusters with different ages. Stars from all 12 clusters are projected toward the traditional USco region, while nine out of the 12 clusters have been assigned to the USco clusters and the remaining three clusters have been assigned to the UCL clusters (see Table~\ref{tab:ages}).   

Figure~\ref{fig:12-hrds-fang17-all} shows all stellar members of the 12 clusters in the \textit{Gaia} BRPR CMD and Fig.~\ref{fig:12-hrds-fang17} displays the individual clusters separately. In these figures, the blue symbols represent all stellar members from the 12 clusters as selected in \citetalias{Ratzenboeck2022}, with additional photometric quality criteria from Eq.\,(\ref{eq:phot-cut}), but excluding the RUWE cut. The red dots represent the sources that are in both samples, \citetalias{Ratzenboeck2022} and \citetalias{Fang2017}. The isochrones in Figs.~\ref{fig:12-hrds-fang17-all} \& \ref{fig:12-hrds-fang17} are PARSEC isochrones for the \textit{Gaia} DR3 passbands with solar metallicity and no extinction. It becomes now clear that previous USco age estimates have been using a mix of different populations at different evolutionary stages and different locations along the line-of-sight. Such a mixture will naturally broaden the HRD or CMD sequence, as is apparent in Fig.~\ref{fig:12-hrds-fang17-all}. 

A possible mixture of populations was already pointed out by \citetalias{Fang2017}, while the available data at that time did not allow a clear separation of the stellar clusters, as was achieved with \textit{Gaia} data. Evidently, separating the \citetalias{Fang2017} sample into coeval clusters, as done by \texttt{SigMA} and other recent studies, solves the age controversy.

\section{Conclusions}\label{sec:conclusion}

In this paper, we reconstruct the star formation history of the closest OB association to Earth, Sco-Cen, by deriving robust isochronal ages for 37 clusters selected with the \texttt{SigMA} algorithm on \textit{Gaia} DR3 data \citep{Ratzenboeck2022}. The ages of the 37 coeval stellar clusters, some previously unrecognized, reveal the complex star formation history of Sco-Cen and are compared with previous work. The main results of this work can be summarized as follows:

   \begin{enumerate}
    
    \item Sco-Cen's star formation history is dominated by a brief period of intense star and cluster formation rate about 15 Myr ago. This is consistent with previous works. Most of Sco-Cen stars and clusters were in place after this intense formation period. The production of stars and clusters has been slowly declining since this burst.

    \item We identified four discernible stages during the formation of Sco-Cen associated with elevated star formation activity. They are, approximately, the $20$\,Myr, $15$\,Myr, $10$\,Myr, and $5$\,Myr bursts. Remarkably, these elevated star formation activity periods seem periodic, separated by spans of about 5\,Myr.  

    \item The formation of stars and clusters is correlated throughout the entire star formation history of Sco-Cen. Still, after the initial burst 20\,Myr ago, the star formation rate more than doubles during the main $15$\,Myr burst. This implies that the formation of the large majority of clusters with supernova precursors (clusters containing more than about 500 stars) took place during the peak of the star- and cluster-formation rate. 

    \item Sco-Cen was formed inside out, meaning that there is a correlation between the age of a cluster and its distance to the oldest cluster in the association. Older clusters from the $20$\,Myr and $15$\,Myr bursts are located in the center of the association, while younger clusters are located toward the outskirts of the association.  

    \item We find well-defined patterns of star formation progression in space and time. In particular, two 100-pc long chains (LCC and CrA chains) of contiguously located clusters exhibit a well-defined age gradient, from massive older clusters to smaller younger ones. The simplest explanation for these long chains of correlated clusters is feedback acting on a diminishing gas reservoir. These patterns are reminiscent of the classic \cite{Elmegreen1977} scenario, suggesting an important role for feedback on the formation of the Sco-Cen population. Morphologically, the formation appears to have been ``Octopus-like'', with most older stars in the head and younger stars in the radial arms, the quasi-linear chains of clusters.  

    \item We confirm the post-\textit{Gaia} view from recent studies that USco is not a single cluster, which solves the Upper-Sco age controversy. What was taken in the literature of the last decades as the USco stellar population consists instead of up to nine clusters with ages between 3 and 19\,Myr, naturally explaining the wide age spread and conflicting results in earlier studies. This realization applies to all Blaauw's subgroups (USco, UCL, and LCC). It directly impacts planet formation studies in Sco-Cen, a benchmark laboratory for planet formation, calling for a revision of disk ages.

  \end{enumerate}

\textit{Gaia} studies of Sco-Cen are revealing a new set of captivating stellar substructures. The classical Blaauw subgroups (USco, UCL, and LCC), originally defined on the plane of the sky in 2D, do not capture the richness of structure and the many stellar populations in Sco-Cen. Separation into three main regions is obsolete and does not encapsulate the more complex, but more revealing star-formation history of this association. Tracebacks of the different Sco-Cen clusters will test the main conclusions of this work, and they will be able to test and characterize the existence of well-defined chains of clusters in OB associations.

\begin{acknowledgements}
    S.\,Ratzenb{\"o}ck acknowledges funding by the Austrian Research Promotion Agency (FFG, \url{https://www.ffg.at/}) under project number FO999892674.
    J.\,Gro\ss schedl acknowledges funding by the Austrian Research Promotion Agency (FFG) under project number 873708. Co-funded by the European Union (ERC, ISM-FLOW, 101055318). Views and opinions expressed are, however, those of the author(s) only and do not necessarily reflect those of the European Union or the European Research Council. Neither the European Union nor the granting authority can be held responsible for them.
    This work has made use of data from the European Space Agency (ESA) mission {\it Gaia} (\url{https://www.cosmos.esa.int/gaia}), processed by the {\it Gaia} Data Processing and Analysis Consortium (DPAC, \url{https://www.cosmos.esa.int/web/gaia/dpac/consortium}). Funding for the DPAC has been provided by national institutions, in particular, the institutions participating in the {\it Gaia} Multilateral Agreement. 
    This research has used Python, \url{https://www.python.org}; \textit{Astropy}, a community-developed core Python package for Astronomy \citep{Astropy2013, Astropy2018}; 
    NumPy \citep{Walt2011}; 
    Matplotlib \citep{Hunter2007};  
    and Plotly \citep{plotly2015}. 
    This research has made use of the SIMBAD database operated at CDS, Strasbourg, France \citep{Wenger2000}; of the VizieR catalog access tool, CDS, Strasbourg, France \citep{Ochsenbein2000}; and of ``Aladin sky atlas'' developed at CDS, Strasbourg Observatory, France \citep{Bonnarel2000, Boch2014}. This research has made use of TOPCAT, an interactive graphical viewer and editor for tabular data \citep{Taylor2005}.
\end{acknowledgements}
\bibliographystyle{aa.bst}
\bibliography{sebastian}

\begin{appendix}

\section{Fitting isochrones to photometric data} \label{apx:method}

We fit model isochrones to the 37 extracted \texttt{SigMA} clusters to infer cluster ages. Isochronal age estimates are an effective tool for studying relative ages and age sequences in the CMD, as consistent shifts in the position of pre--main-sequence (PMS) objects relative to the main-sequence (MS), precise photometry, and parallax measurements from \textit{Gaia} provide reliable evidence for relative age differences between populations. These systematic shifts are especially pronounced in young populations ($\leq 100$\,Myr). As older populations reach the main sequence, the changes in the isochrones below the turn-off become almost imperceptible due to the long-lasting, stable process of hydrogen burning, which keeps the CMD distribution virtually stagnant \citep[][]{2010Soderblom}. 

Although relative age sequences become apparent when comparing young populations in the CMD, determining absolute ages is a process that is prone to biasing effects of systematic uncertainties. Such uncertainties, for instance, arise from the use of different isochronal model families~\citep[see e.g.,][]{Kerr2022}. To mitigate the systematic effect of different evolutionary models on our method, we employ two model families: PARSEC and BHAC15 (Sect.~\ref{sec:data}). 

Another systematic uncertainty is introduced by the choice of the fitting technique itself, as different fitting approaches are accompanied by disparate assumptions about the underlying data distribution of the measurements. For example, the popular and often used least squares (LS) method is based on the expectation that the data exhibit Gaussian uncertainties around the regression curve. 
However, assuming a Gaussian distribution of the data may not be entirely justified in the first place, as it is well-established that observational trends such as the unresolved binary sequence appear brighter compared to the single-star main sequence. At the same time, extinction skews the source distribution towards fainter magnitudes and redder colors. As a result, a fit of the (bluer) left-main ridge of the data in the CMD is often the preferred location for an isochrone, as it is believed to represent the underlying data more accurately. This can, for example, be achieved with fitting techniques that do not implicitly assume Gaussian distribution (see Sect.~\ref{sec:stat_model}).

As susceptibility to outliers is a general problem (and not specifically related to LS), we aim to curb their influence by applying photometric quality filters (see Sect.~\ref{sec:data} for a detailed outline of the applied quality criteria) and employing robust fitting techniques. Robust fitting techniques are less sensitive to the influence of outliers in the data~\citep{Rousseeuw:2005}. As we still observed outliers in the CMD distributions after applying our quality cuts, we aim to use techniques that decrease their influence, producing more stable and reliable results. In the following, we briefly discuss robust methods and the choice of an appropriate method depending on the specific characteristics of the data.

\begin{table*}[!t] 
\begin{center}
\begin{small}
\caption{Description of the parameters and flat prior ranges.\vspace{-2.5mm}}
\renewcommand{\arraystretch}{1.3}
\resizebox{1.95\columnwidth}{!}{
\begin{tabular}{lcccl} 
\hline \hline

\multicolumn{1}{c}{} &
\multicolumn{2}{c}{Parameter} &
\multicolumn{1}{c}{Range} &
\multicolumn{1}{l}{Description} \\ 

\cmidrule(lr){2-3}

\multicolumn{1}{c}{} &
\multicolumn{1}{c}{Model} &
\multicolumn{1}{c}{Astrophysical} & 
\multicolumn{1}{c}{} &
\multicolumn{1}{l}{} \\ 

\cmidrule(lr){1-5}

Isochrone parameters & $\tau$ & logAge & (6, 8) & Cluster age (log$_{10}$ yr) \\
 & $\zeta$ & [Fe/H] & 0 & Cluster metallicity (dex) \\
 & $\epsilon$ & $A_V$ & [0, 2] & Dust extinction in V band (mag)\\

\cmidrule(lr){1-5}

Skewed Cauchy PDF parameters\tablefootmark{a} & $s$  ($G_\mathrm{BP} - G_\mathrm{RP}$) & -- & (0.15, 0.9) & Scale parameter determined using $G_\mathrm{BP} - G_\mathrm{RP}$ colors \\
& $s$ ($G - G_\mathrm{RP}$) & -- & (0.03, 0.1) & Scale parameter determined using $G - G_\mathrm{RP}$ colors \\  
& $a$ & -- & (0.15, 0.9) & Skewness, constant across photometric systems\\




\hline
\end{tabular}
} 
\renewcommand{\arraystretch}{1}
\label{tab:parameters_priors}
\tablefoot{
\tablefoottext{a}{The PDF parameter ranges have been determined via maximum likelihood fits of Eq.~(\ref{eq:cauchy_pdf}) to data and isochrones from studies by~\citet{Bossini2019} and~\citet{Dias2019}. We limit the sample to only include clusters within 500\,pc and with ages less than 100\,Myr, to guarantee similar CMD conditions compared to the Sco-Cen subgroups.}
}
\end{small}
\end{center}
\end{table*}

\subsection{Robust fitting methods}

Among the different robust methods available, two prominent approaches to robust fitting include employing non-parametric methods, which aim to reduce the impact of outliers on observations that otherwise have an underlying normal distribution, and using a (more) robust loss function for regression. Which of these two options is generally better depends on the specific context and objectives of the scientific analysis.

Non-parametric methods for robust regression are a class of statistical techniques that do not make any assumptions about the underlying distribution of the data. These methods are typically less sensitive to outliers and are often employed when the underlying distribution is unknown. Commonly used methods include Least Trimmed Squares~\citep[LTS,][]{Rousseeuw:1984} and RANdom SAmple Consensus~\citep[RANSAC,][]{Fischler:1981}. LTS works by iteratively removing a pre-defined fraction of the data that exhibit the largest residuals and then minimizing the sum of squared residuals over the remaining subset. The RANSAC algorithm works by iteratively selecting and fitting random subsets of the data and determining their outlier fraction. The best fit minimizes the number of outliers. The subset size and the outlier threshold are critical parameters of LTS and RANSAC, respectively, which depend on the inherent fraction of outliers in the data and their approximate distance from the remaining ``inliers''. If not set carefully, they can affect the resulting model and lead to poor results.

Robust loss functions are less sensitive to outliers than the traditionally used normal distribution due to their utilization of heavy-tailed distributions. Their heavy tail effectively assigns a smaller weight to data points that deviate significantly from the overall trend, decreasing the influence of outliers. Prominent candidates include the Huber loss~\citep{Huber:1964} and the Tukey biweight loss~\citep{Beaton:1974}. Both methods combine the mean squared error loss with a lesser penalizing loss for outlying observations. The difference between respective loss functions is that the Huber loss scales linearly with outliers above a given threshold. In contrast, Tukey's loss is even less sensitive to outliers, as it flattens to a constant value at the given outlier threshold.

Generally, both non-parametric and parametric regression techniques try to reduce the impact of outliers while simultaneously fitting a least-sum of squares model to normal observations. In other words, they assume a symmetric data distribution of inliers around the ``true'' model. As mentioned, CMD data is known to deviate from this assumption and show a skewed distribution (towards fainter and cooler stars) of sources due to the appearance of reddening and unresolved objects in the data. To solve this predicament, we chose to approximate the data distribution with a skewed Cauchy likelihood function to capture the source distribution around the isochrone better.

\subsection{Statistical model}\label{sec:stat_model}

Applying quality filters (see Sect.~\ref{sec:data}) drastically reduces the data size and range of photometric uncertainties (see also Fig.~\ref{fig:hrd-all}). We generally do not observe strong heteroscedasticity within the sample. On the contrary, in the case of outliers, the reported uncertainties are often too inconsequential to explain large observed deviations from theoretical isochrones. As mentioned, observational data may be biased by reddening or resolution limits. Moreover, cluster selections could be influenced by contamination from older sources, which, however, is generally low for the used sample ($\sim$6\%), as determined in \citetalias{Ratzenboeck2022}. Still, even more effects, such as small intra-cluster metallicity variations or stellar rotation, can result in a departure from the regression curve of a given model. We are generally unable to determine the exact nature of each data point's deviation, and we have to account for all these effects in our choice of model as best as possible.

The skewed Cauchy distribution extends the standard Cauchy distribution, allowing for skewness. The standard Cauchy distribution is a heavy-tailed distribution known for its robustness to outliers and ability to model data with many outliers~\citep{Hampel:2011}. The PDF of the zero-centered skewed Cauchy distribution is defined by a scale parameter $s$, with $s \in \mathbb{R}^{+}$, which controls the width of the distribution, and a skewness parameter $a$, with $a \in (-1, 1)$:
\begin{equation}\label{eq:cauchy_pdf}
    f(x; s, a) = \left[s\pi \left( 1 + \frac{x^2}{s^2 (1 + a\, sign(x))^2} \right) \right]^{-1}
\end{equation}
Setting $a = 0$ yields the standard Cauchy distribution.

\subsubsection{Likelihood}
For a given set of model parameters $\Vec{\theta} = (\tau, \zeta, \epsilon)$ denoting age, metallicity, and dust extinction, respectively, the stellar evolution model predicts a set of magnitudes that characterize a co-eval single-stellar
population\footnote{Stellar evolution models are also parameterized by the initial mass of a star. We consider an isochrone model the curve resulting from varying the mass between $0.1 \leq M/M_{\odot} < 50$.}. The astrophysical parameter equivalent to the model parameter vector is (logAge, [Fe/H], $A_V$). It describes age in log$_{10}$ years, metallicity in log$_{10}$ of the iron to hydrogen ratio (calibrated to the sun), and extinction measured in the V band in magnitudes, see Table~\ref{tab:parameters_priors}.

To describe the observations, we assume a simple model in which the data $\{D_n\}_{n=1}^{N}$ (of size $N$) is generated along isochrones with noise contributions $e_n$ drawn independently from skewed Cauchy distributions with zero means. We consider the models in the CMD as tuples of colors $c$ and magnitudes $m$. The expected color and magnitude of an observed star $n$, conditioned on model parameters $\Vec{\theta}$, is determined by the theoretical point on the isochrone\footnote{Isochronal models are available on a grid in age and metallicity ($\Delta \tau = 0.2$, see Table~\ref{tab:parameters_priors} for units), which we interpolate linearly between equal-mass grid points. For a given age-metallicity tuple, we move the isochrone in the CMD based on a given dust extinction value $\epsilon$, assuming a constant extinction law for each \textit{Gaia} filter.}, which minimizes the distance to the observed star's color $c_n$ and magnitude $m_n$. Their signed distance $d(\Vec{\theta})$ is defined as follows:
\begin{equation}
    d_n(\Vec{\theta})^2 = (c_n - c(\Vec{\theta}))^2 + (m_n - m(\Vec{\theta}))^2
\end{equation}
The sign of $d(\Vec{\theta})$ is determined according to whether the observed star is redder or bluer than the isochrone. 

For simplicity, we assume that every data point is drawn from the same constant-noise model. Further, we treat non-symmetric noise sources such as unresolved binaries and reddening effects as mass-independent. Thus, we employ a constant scale parameter $s$. 
Depending on the location and distribution of sources relative to the interstellar medium (ISM), reddening effects can vary widely among clusters, which translates to different scales $s$ of the skewed Cauchy distribution\footnote{To a first approximation, the scale parameter $s$ handles differential reddening effects. This allows us to work with a single extinction value, although more complex reddening effects might exist. In practice, the inference finds approximately the minimum extinction value throughout the cluster, while the scale parameter $s$ takes higher differential reddening into account (to some degree).}. Instead of fixing $s$, it becomes a free parameter of the model, and its value is obtained during Bayesian inference. Similarly, the skewness parameter $a$ is determined during inference. The distribution $p(D_n \mid \Vec{\theta}, s, a)$ of the data points $D_n$ becomes the following:
\begin{equation}
    e_n \sim p(D_n \mid \Vec{\theta}, s, a) = f(d_n(\Vec{\theta}); s, a)
\end{equation}
By combining the isochrone parameters $\Vec{\theta}$ and the noise distribution parameters $(s, a)$ into $\Theta = (\tau, \zeta, \epsilon, s, a)$, we can write the likelihood as:
\begin{equation}
    \mathcal{L}(\{D_n\}_{n=1}^N \mid \Theta) = \prod_{n=1}^{N} p(D_n \mid \Theta)
\end{equation}

\subsubsection{Posterior probability}\label{sec:posterior}

Applying Bayes' theorem, we obtain the posterior PDF of the parameters $\Theta$, conditional on the data $\{D_n\}_{n=1}^N$:
\begin{equation}\label{eq:posterior}
    p(\Theta \mid \{D_n\}_{n=1}^N) \propto  \mathcal{L}(\{D_n\}_{n=1}^N \mid \Theta) ~ p(\Theta)
\end{equation}
The PDF $p(\Theta)$ represents our prior knowledge of the parameters $\Theta$, which we summarize in Table~\ref{tab:parameters_priors}. We employ flat priors and limit the isochrone parameters to ages between 1--100\,Myr, dust extinction in the V band to $A_V \in [0,2]$, and set the metallicity to [Fe/H] = 0 (solar metallicity). Due to the high degeneracy between age and metallicity, fixing the metallicity to the solar value reduces ambiguity in age. However, one must be careful when omitting the metallicity parameter from the fit due to its impact on absolute isochronal ages. In the solar neighborhood, assuming solar metallicity is a common approximation appropriate for Sco-Cen~\citep{VianaAlmeida:2009} and consistent with the short mixing timescales measured in supernova-driven ISM simulations~\citep{deAvillez:2002}.

The prior ranges of PDF parameters are established through maximum likelihood fits of Eq.\,(\ref{eq:cauchy_pdf}) to data around best-fitting isochrones found by \citet{Bossini2019} and~\citet{Dias2019}. To ensure comparability, we apply the same quality criteria discussed in Sect.~\ref{sec:data} and restrict the sample to clusters within 500\,pc and to ages $<$~100\,Myr to create similar CMD conditions as Sco-Cen sub-clusters. We assume flat priors between the minimum and maximum of the obtained parameter values. Since different colors have different value ranges, we determine separate scale ranges for $G_\mathrm{BP} - G_\mathrm{RP}$ and $G - G_\mathrm{RP}$. 

We use the Markov Chain Monte Carlo (MCMC) method implemented within the public code \texttt{emcee}~\citep{ForemanMackey:2013} to generate samples from the posterior PDF in Eq.\,(\ref{eq:posterior}). For each parameter, we compute the marginal PDF, its maximum a posteriori (MAP) position, and the $1\sigma$ credible interval determined via computing the $68$\% high-density interval (HDI) to represent the fitting results and uncertainties, respectively.

\subsection{Validation} \label{sec:validation}

\begin{figure}[!t]
    \centering
    \includegraphics[width=0.88\columnwidth]{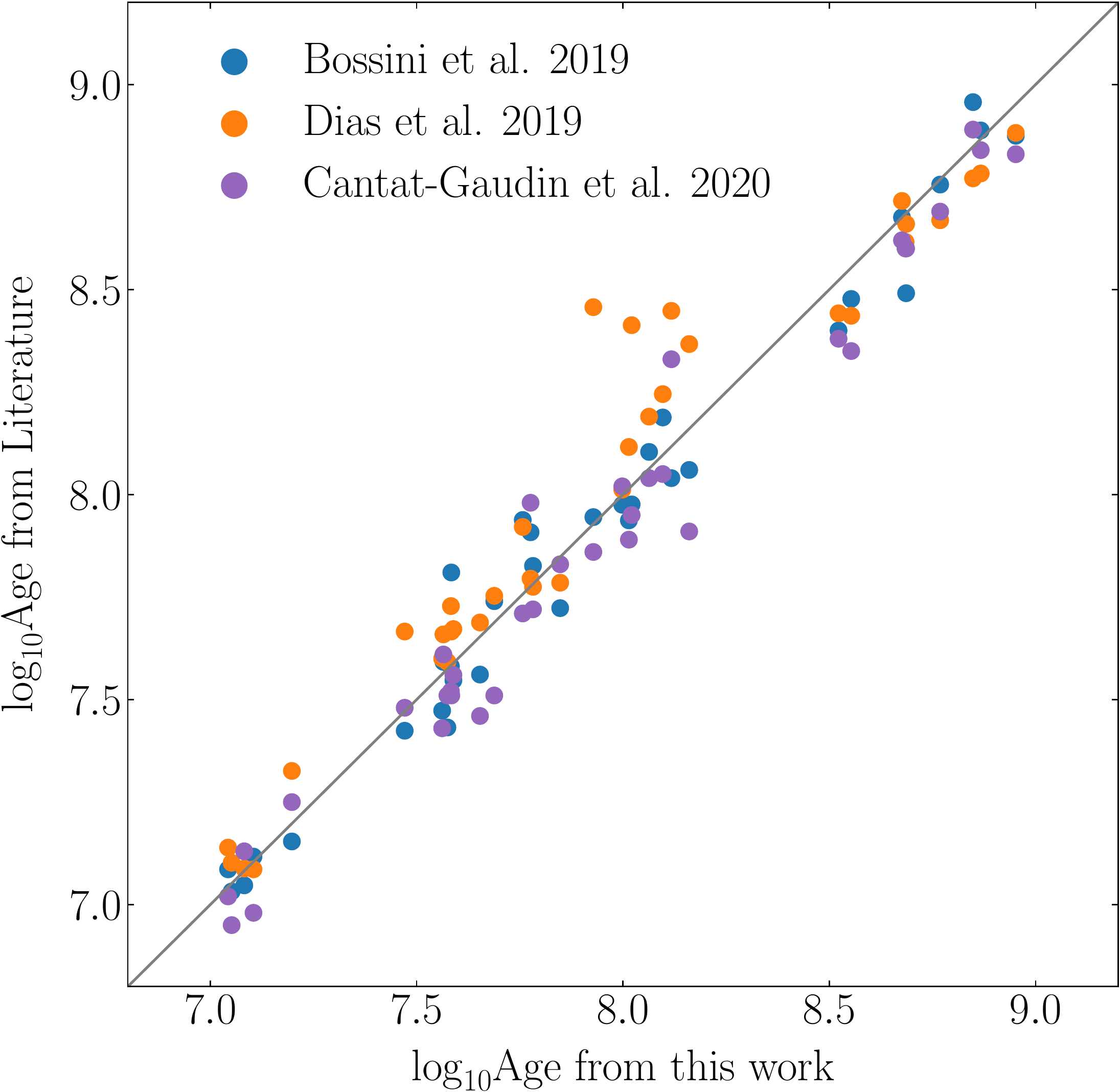}
    \caption{Comparison between our age estimates and ages determined in two previous studies for 35 nearby ($<500$\,pc) open clusters (see legend).}
    \label{fig:validation_literature}
\end{figure}

\begin{figure}[!t]
    \centering
    \includegraphics[width=0.95\columnwidth]{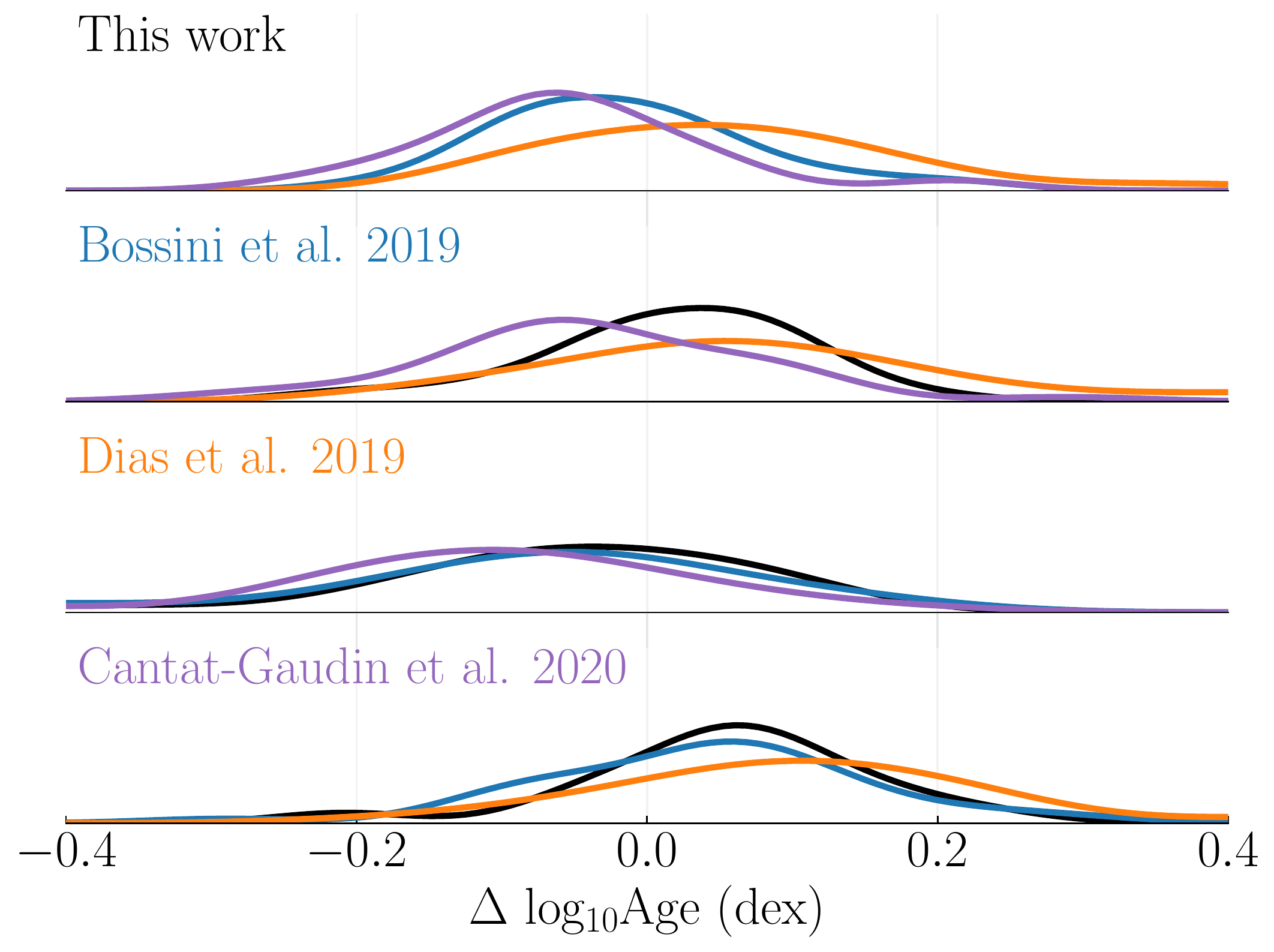}
    \caption{Distribution of logAge differences between this study (in black) and the studies by \citet{Bossini2019} (in blue), \citet{Dias2019} (in orange), and \citet{Cantat-Gaudin2020b} (in purple). The distributions are obtained via a KDE, using Scott's rule~\citep{Scott:1979} to determine the bandwidth. The bottom three panels show the distribution of logAge differences for the remaining surveys against the other three, respectively. Using this comparison, we find no systematic differences in estimated ages across the four presented methods.}
    \label{fig:validation_comparison}
\end{figure}

To validate our methodology, we selected a subset of clusters from \cite{Cantat-Gaudin2020a} that are within 500\,pc and with more than 100 members. This selection results in 35 clusters, which have ages determined by \citet{Bossini2019}, \citet{Dias2019}, and \citet{Cantat-Gaudin2020b} in the age range 7$<$\,logAge\,$<$9. We apply our age fitting to the same data using the PARSEC models and the BPRP CMD, also used by the mentioned literature.
The differences between the ages determined in this work and these studies are consistently within the boundaries of one standard deviation (see Fig.~\ref{fig:validation_literature}). The differences in logAge have an average of 0.02\,dex, and a standard deviation of 0.09\,dex when compared to \citet{Bossini2019}. Compared to \citet{Dias2019}, the mean deviation is -0.07\,dex, and the standard deviation is 0.14\,dex. For \citet{Cantat-Gaudin2020b}, the mean difference is 0.05\,dex, and the standard deviation is 0.1\,dex. The top panel of Fig.~\ref{fig:validation_comparison} displays the distribution of \mbox{logAge} differences between this study and the studies by \citet{Bossini2019}, \citet{Dias2019}, and \citet{Cantat-Gaudin2020b}. These are represented by blue, orange, and purple colors, respectively. We have used a KDE\footnote{We determined the bandwidth by employing Scott's rule~\citep{Scott:1979}.} to approximate the shape of the distribution. The bottom three panels show the distribution of logAge differences for the remaining surveys against the other three, respectively. We find similar uncertainties across all surveys; thus, there seem to be no significant systematic differences across these four age estimates.

\begin{figure}[!t]
    \centering
    \includegraphics[width=0.97\columnwidth]{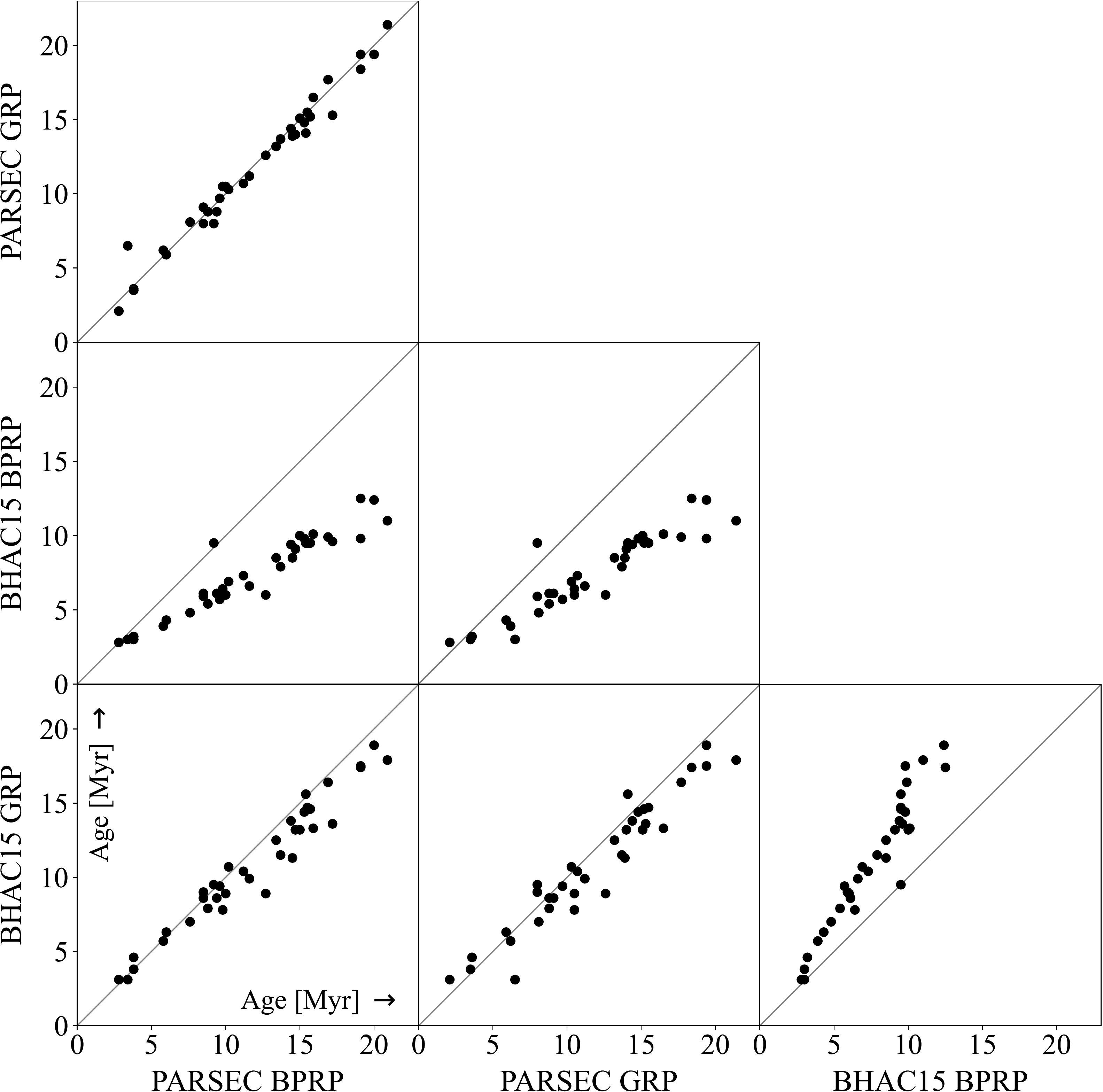}
    \caption{Comparison of the cluster ages (in Myr) as determined with different model families (PARSEC and BHAC15) and different \textit{Gaia} CMDs (BPRP and GRP). We observe strong agreement between ages determined with PARSEC BPRP, PARSEC GRP, and BHAC15 GRP isochrones. In contrast, BHAC15 BPRP model fits seem to suggest systematically younger ages.}
    \label{fig:age_splom}
\end{figure}

\subsection{Age dependency on model families and photometric systems} \label{sec:age-comparison}

In this section, we discuss the sources of systematic uncertainty between determined ages as introduced by different model isochrone families and photometric systems. 
For a detailed summary of the inferred ages, see Table~\ref{tab:ages}. It also includes 1$\sigma$ confidence intervals determined via the highest density interval (HDI) from the marginalized posterior PDF.

Figure~\ref{fig:age_splom} compares the four determined cluster ages. We find good agreement between different color spaces (BPRP, GRP) when employing PARSEC isochrones with systematic uncertainties in the order of $\lesssim 1$\,Myr. Different isochrone model families (PARSEC, BHAC15) also agree well when using GRP colors. Comparing BHAC15-GRP to the PARSEC solutions, we find that older populations tend to be about $1.8 \pm 1.6$\,Myr younger when estimated with the BHAC15-GRP. At the same time, we do not identify a bias for young populations (<10\,Myr) in the inferred age (estimated ages agree within $\lesssim 1$\,Myr). The most significant systematic trend is introduced when estimating ages with BHAC15 models in BPRP colors. We identify an approximately linear trend between BHAC15-BPRP ages and the other model families and color spaces, which leads to an underestimation of the ages by roughly $60$\%, compared to the other results.

\section{Comparison to Kerr et al. (2021)} \label{apx:kerr}

In \citetalias{Ratzenboeck2022}, we compare the \texttt{SigMA} clustering results in more detail with the results of \citet{Kerr2021} (hereafter, \citetalias{Kerr2021}), who use HDBSCAN to select young clusters within 333\,pc from the Sun, including the Sco-Cen association. We find a similar extent and clustered substructure in the Sco-Cen region. However, the individual \texttt{SigMA} clusters are significantly larger in size and numbers of stars compared to \citetalias{Kerr2021} \citepalias[see Table\,E.3 in][]{Ratzenboeck2022}.

\begin{figure}[!t]
        \centering
        \includegraphics[width=0.95\columnwidth]{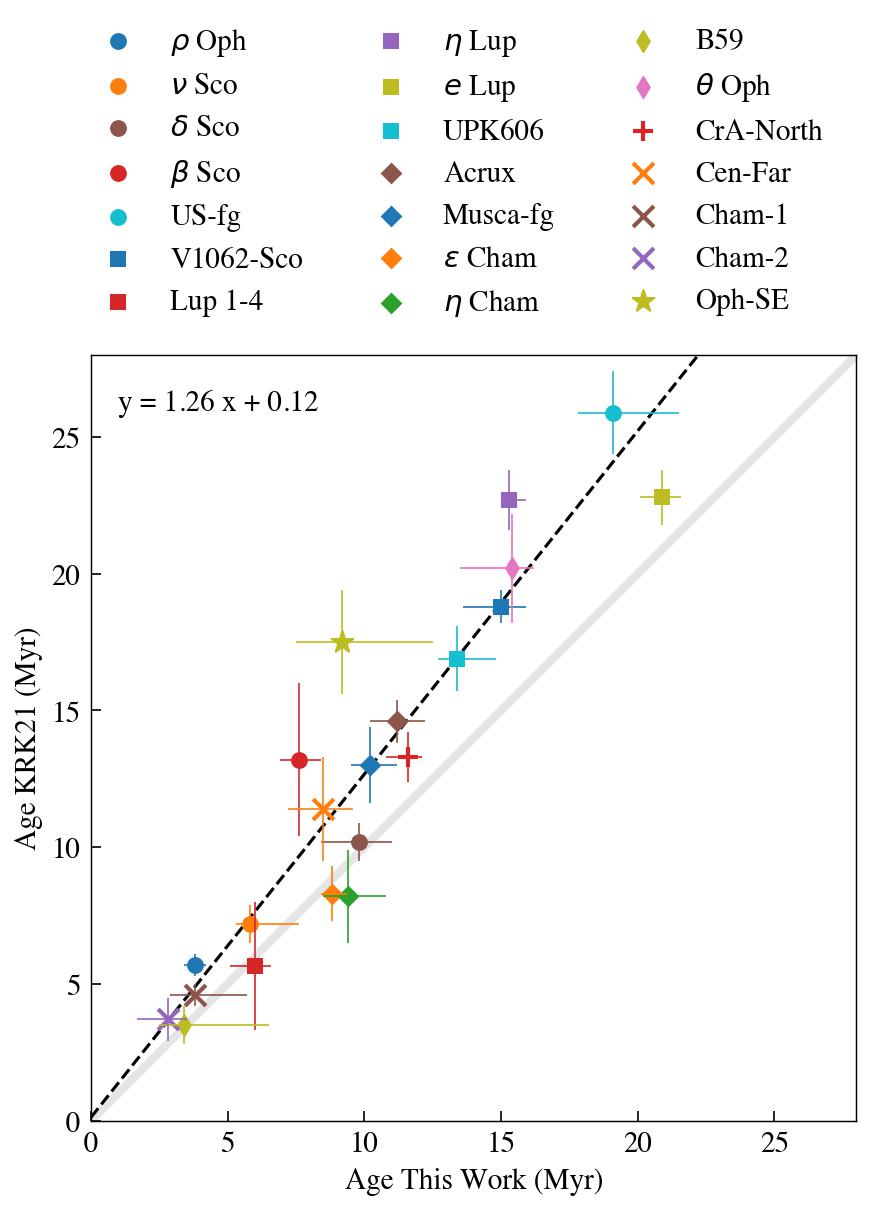}
        \caption{Comparison of the cluster ages from this work (using PARSEC-BPRP) to ages from \citetalias{Kerr2021}. Only clusters with sufficient overlap between the two samples are shown (see text for more information). The solid gray line is a one-to-one line, and the black dashed line is a linear fit to the data points, as given in the panel. Individual clusters are marked with different colors and symbols (see legend), including error bars.}
        \label{fig:comparing_kerr_individual}
    \end{figure}

Figure~\ref{fig:comparing_kerr_individual} compares the age estimates from \citetalias{Kerr2021} to ages as determined in this work using the PARSEC-BPRP ages if sufficient overlap between individual clusters is present. \citetalias{Kerr2021} also uses the PARSEC-BPRP models with solar metallicity, allowing direct comparison. 
We require that at least 10\% of the sources of a matching \texttt{SigMA} cluster need to be part of the corresponding \citetalias{Kerr2021} cluster and that at least 60\% of the sources of one \citetalias{Kerr2021} cluster need to be part of the same \texttt{SigMA} cluster. The different fractions are chosen since the individual \citetalias{Kerr2021} clusters are significantly smaller in size compared to \texttt{SigMA}, in particular when considering sub-clusterings which are themselves parts of \citetalias{Kerr2021}'s top-level clusterings (TLC) to extract more substructure (sub-clustered by \citetalias{Kerr2021} with HDBSCAN's EOM or \textit{leaf}).  

In Figure\,\ref{fig:comparing_kerr_individual} there appears to be a trend with growing age in that \citetalias{Kerr2021} ages are systematically older as a function of our ages, affecting in particular clusters with older ages ($\gtrsim 10$\,Myr). 
This trend is puzzling because both studies use the same isochrone models. The imperfect overlap between the two samples could create different outcomes in the age fitting; however, possible imperfect matches of clusters would not create a trend. Another difference is the use of \textit{Gaia} DR2 in \citetalias{Kerr2021} versus \textit{Gaia} DR3 in \citetalias{Ratzenboeck2022}. 
A likely culprit for the age-difference trend seems to be the age correction done in \citetalias{Kerr2021}, which might be biasing their age estimate as a function of age (with their correction procedure, they will necessarily get more field stars for the older clusters)\footnote{They use the selected clusters as ``signposts'' (training sets) to select additional potential cluster members with HDBSCAN, with similar spatial and kinematic properties, to reintroduce potentially older members, older than their original age selection of < 50\,Myr.}. Moreover, as outlined in the methods section, selecting an appropriate fitting method is crucial since the scatter of sources in a CMD is not distributed normally around the best fitting isochrone (see the explanations in Appendix~\ref{apx:method}).

The age-difference trend highlights the importance of careful age determination when using isochronal models. It should caution against comparing ages from different works at face value without considering possible biases that the various methods and fitting approaches could introduce. 

\section{Additional Figures} \label{aux:fig}

\begin{figure*}
    \centering
    \includegraphics[width=0.97\textwidth]{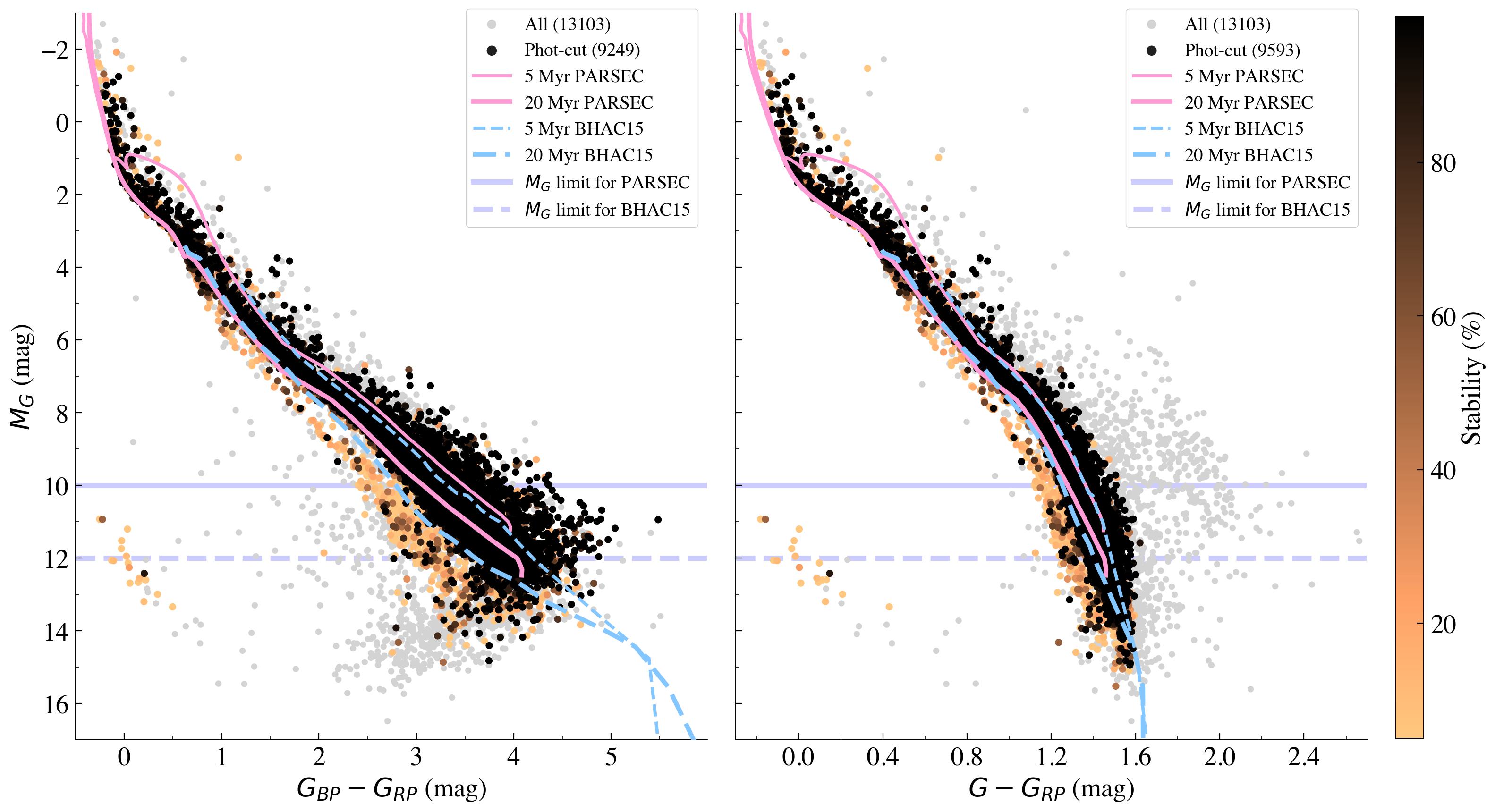}
    \caption{\textit{Gaia} CMDs for the BPRP (left) and GRP (right) colors. The gray sources are all Sco-Cen members from \citetalias{Ratzenboeck2022}. Sources that remain after applied quality criteria (see Sect.\,\ref{sec:data}) are color-coded in copper for stability. The scatter of the gray sources compared to the colored sources highlights the influence of photometric uncertainties. PARSEC and BHAC15 isochrones are over-plotted for 5\,Myr and 20\,Myr. We mark the two magnitude limits at $M_G>10$\,mag and $M_G>12$\,mag in light-purple as used for PARSEC and BHAC15 isochrone fitting, respectively.}
    \label{fig:hrd-all}
\end{figure*}

Figure~\ref{fig:hrd-all} displays the members of all 37 \texttt{SigMA} clusters in the BRPB and GRP CMDs (lightgray dots). After applying the quality criteria from Sect.~\ref{sec:data}, the scatter of sources in the CMD reduces (colored dots), particularly affecting low-mass sources. There is a trend such that inferior photometry tends to get shifted to the left in the BPRP CMD and to the right in the GRP CMD (see the scatter of the lightgray dots). The reliability of a source's cluster membership, as selected with \texttt{SigMA}, can be estimated via a stability value ranging from 0\%--100\%, which indicates how often individual sources appear throughout the ensemble of clustering solutions per cluster.
We color-code the sources by their stability value. It can be seen that sources, which appear on an older age sequence, generally have lower stability and are, therefore, more unreliable members. In this work, we do not remove sources based on a stability cut since our isochrone fitting method is tuned to deal with outliers. 

\begin{figure*}
        \centering
        \includegraphics[width=0.98\textwidth]{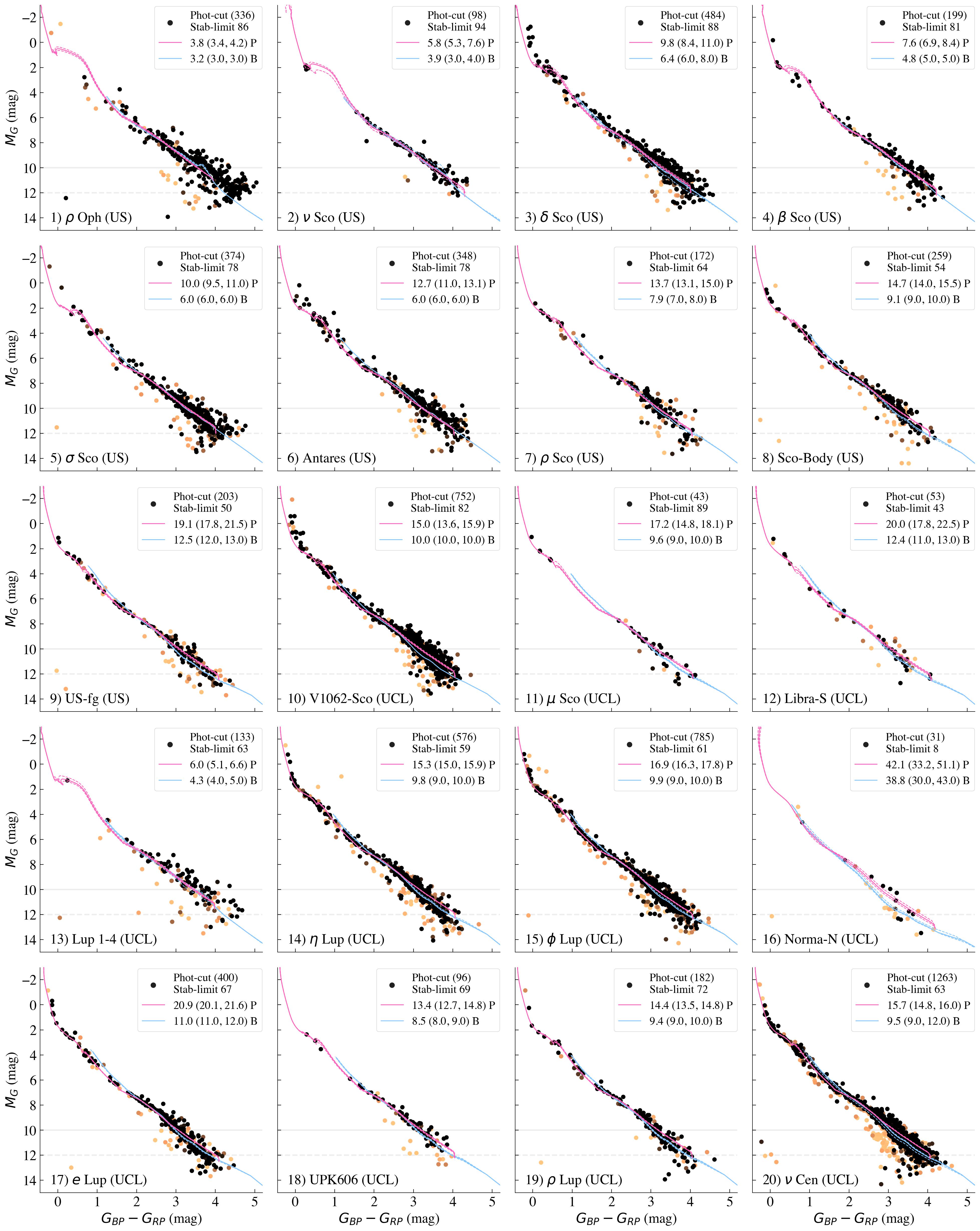}
        \caption{\textit{Gaia} BPRP CMDs for the \texttt{SigMA} clusters 1--20. 
        Only cluster members, which pass the photometric quality criteria from Eq.\,(\ref{eq:phot-cut}) are plotted, with the remaining number of sources given in the legend (Phot-cut). The dots are color-coded for stability with lower limits set to 2.5\% (orange) and upper limits (dark) are varied per cluster, as given in the legend (Stab-limit, \%). The magenta and blue solid lines show the best-fitting PARSEC (P) and BHAC15 (B) isochrones, respectively, as determined for BPRP, and the dashed lines show the upper and lower age limits (age limits are given in parenthesis in the legend). The horizontal solid and dashed light-gray lines give the magnitude limits at $M_G>10$\,mag and $M_G>12$\,mag for PARSEC and BHAC15, respectively, used to exclude sources from the age fitting.}
        \label{fig:BPRP-HRDs-1}
    \end{figure*}

\begin{figure*}
        \centering
        \includegraphics[width=0.98\textwidth]{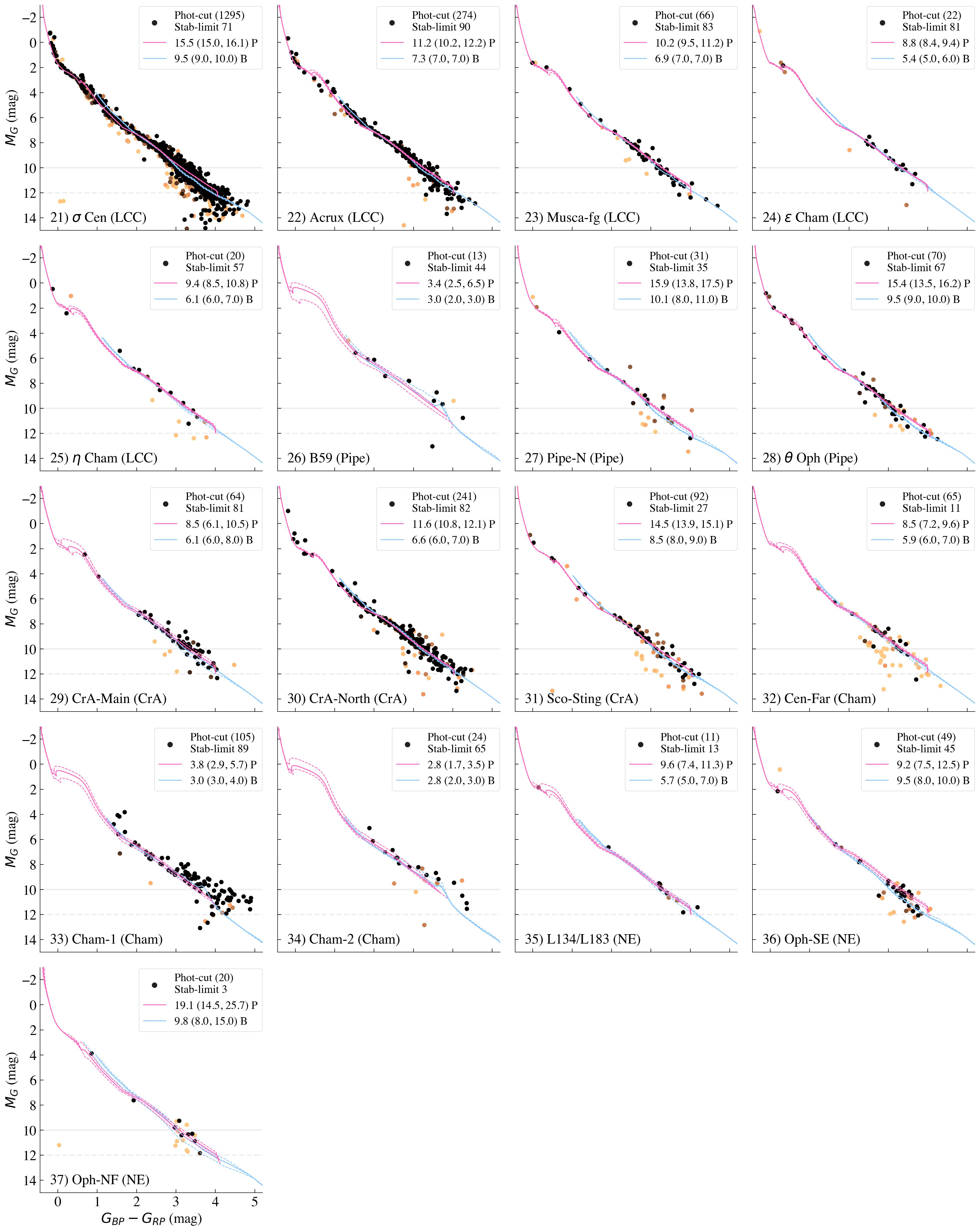}
        \caption{Same as Fig.\,\ref{fig:BPRP-HRDs-1}, for the \texttt{SigMA} clusters 21--37.}
        \label{fig:BPRP-HRDs-2}
    \end{figure*}

\begin{figure*}
        \centering
        \includegraphics[width=0.98\textwidth]{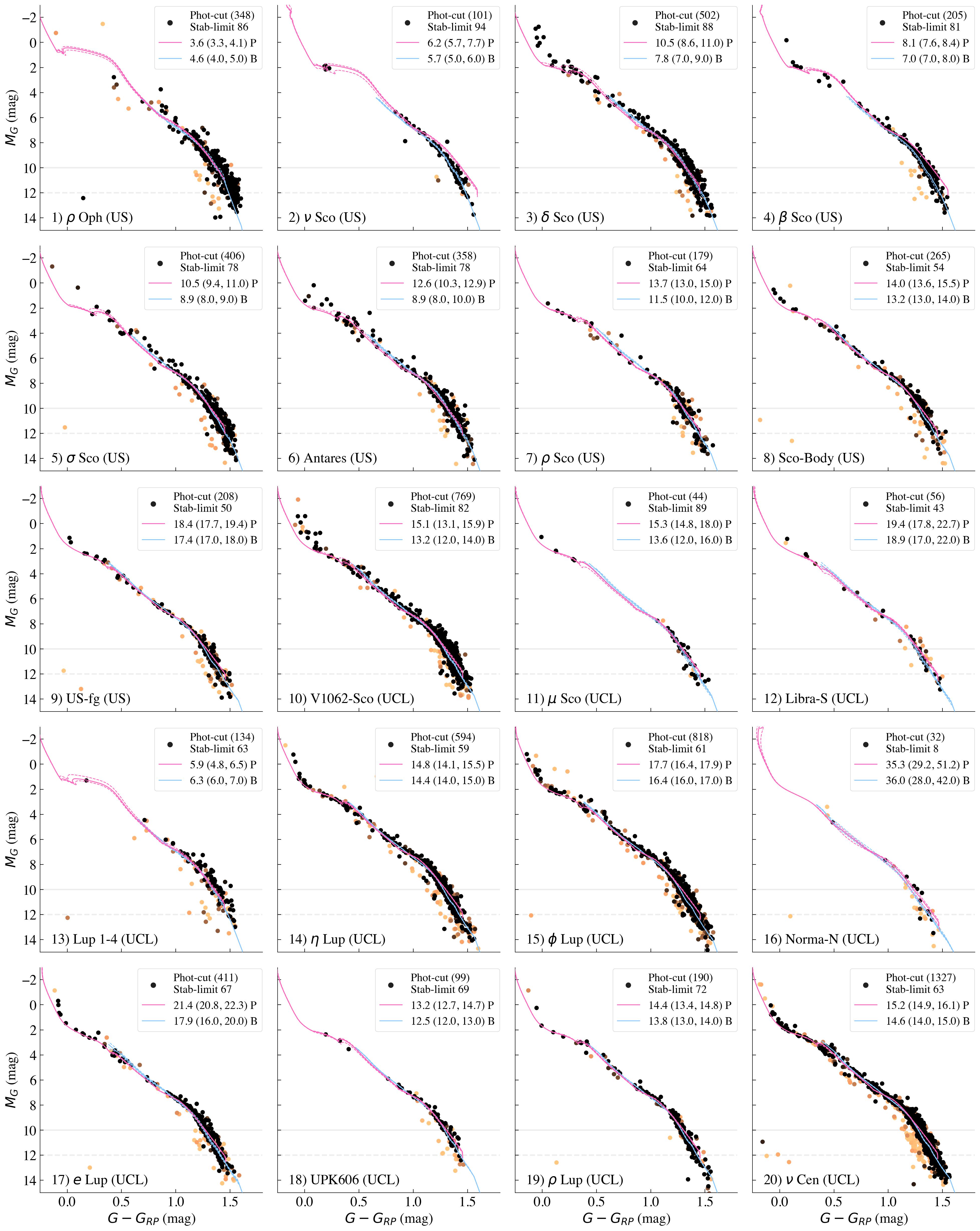}
        \caption{Similar as Fig.\,\ref{fig:BPRP-HRDs-1}, but showing the GRP CMD for the \texttt{SigMA} clusters 1--20. The best fitting PARSEC isochrone is shown as determined with GRP.}
        \label{fig:GRP-HRDs-1}
    \end{figure*}

\begin{figure*}
        \centering
        \includegraphics[width=0.98\textwidth]{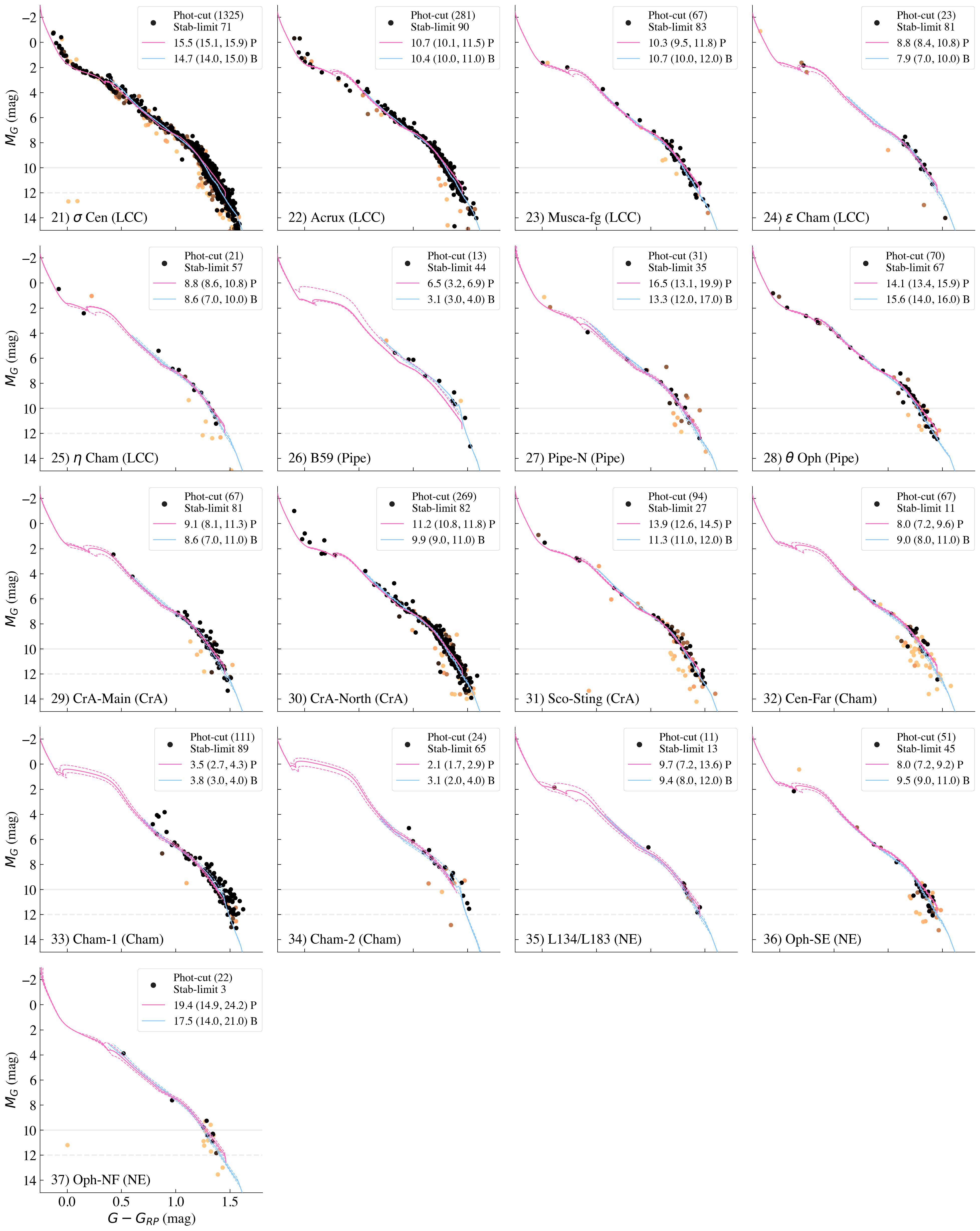}
        \caption{Same as Fig.\,\ref{fig:GRP-HRDs-1}, for the \texttt{SigMA} clusters 21--37.}
        \label{fig:GRP-HRDs-2}
    \end{figure*}

Figures~\ref{fig:BPRP-HRDs-1}--\ref{fig:GRP-HRDs-2} show the \textit{Gaia} CMDs for each cluster with over-plotted isochrones from the best fitting PARSEC and BHAC15 models, both for BPRP and GRP CMDs. The sources are again color-coded for stability. The maximum stability varies per cluster, while the stellar members of the more massive clusters tend to have higher stability up to 100\%, while some smaller scale clusters have a maximum only at around 10\%, like Cen-Far or Oph-NF. This does not indicate that such clusters are not real, while their identification in the sea of noise was less pronounced compared to more massive clusters. Therefore, we vary the upper limits of the color scale individually per cluster, using the mean stability per cluster, which is given in the legend of each panel. 

\begin{figure*}
        \centering
        \includegraphics[width=1.9\columnwidth]{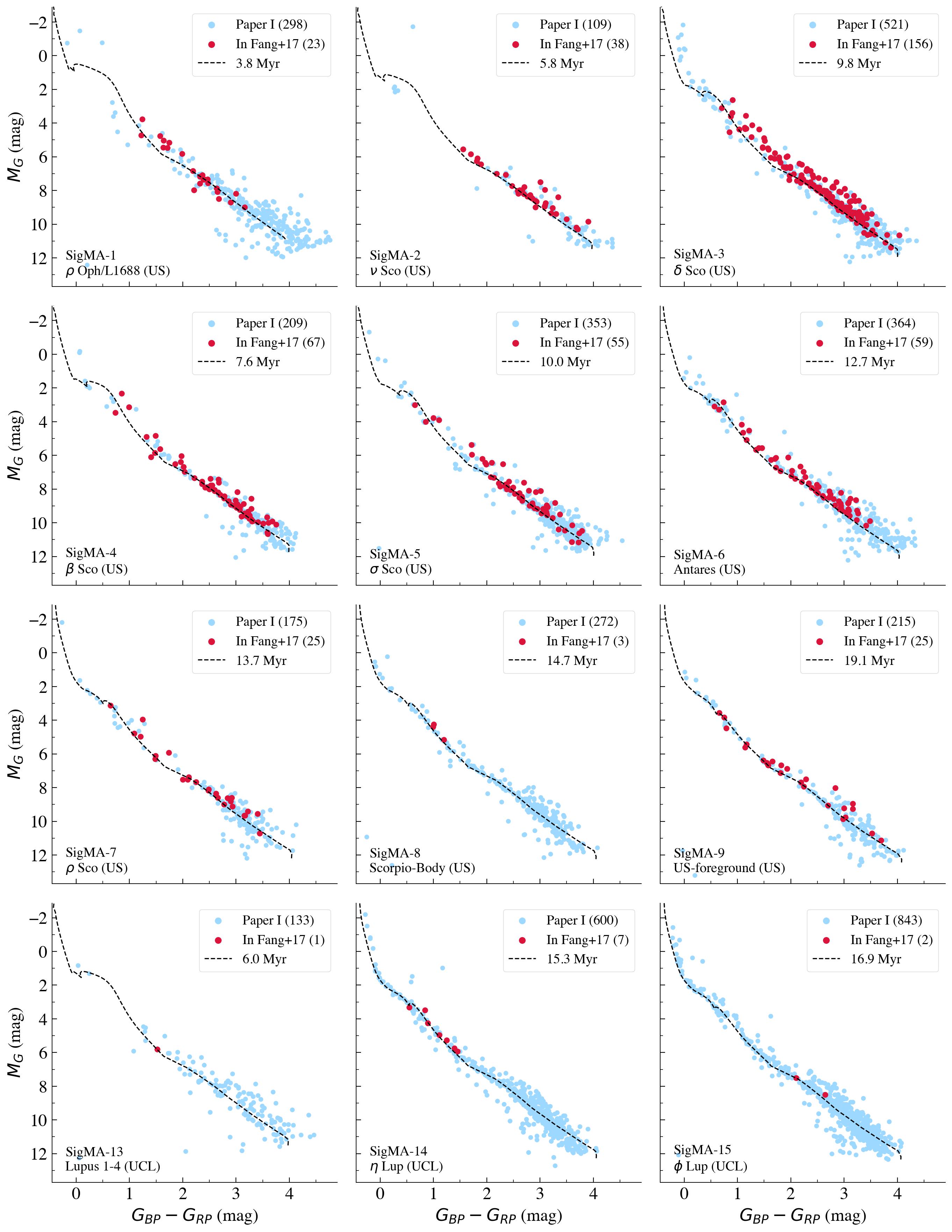}
        \caption{\textit{Gaia} BPRP CMD showing the 12 \texttt{SigMA} clusters that have matches with sources from \citetalias{Fang2017}. Similar to Fig.~\ref{fig:12-hrds-fang17-all}, but now the individual clusters are displayed in separate panels. The sample includes nine clusters from USco and three clusters from UCL (bottom row), as assigned based on traditional borders \citepalias[see][]{Ratzenboeck2022}. The three UCL clusters have only a few matches since they only partially reach into the USco region. The blue dots are the \texttt{SigMA} members of the respective clusters with additional photometric quality criteria (see Sect.~\ref{sec:fang}). The red dots mark the sources that match with the \citetalias{Fang2017} sample. We do not use the RUWE cut for this overview, which reveals some binary sequences. The dashed black line shows the PARSEC isochrone for the cluster age as estimated in this work (see legends).}
        \label{fig:12-hrds-fang17}
    \end{figure*}

Figure~\ref{fig:12-hrds-fang17} 
shows the 12 clusters, which have matches with the USco sample from \citet{Fang2017} (see Sect.~\ref{sec:fang}), first shown in Fig.~\ref{fig:12-hrds-fang17-all}. The separation of the clusters into individual CMDs highlights the different ages of stellar clusters that are located toward USco, which were often treated as one population in the past.

\end{appendix}
\end{document}